\documentclass[pdflatex,sn-mathphys-num]{sn-jnl}

\usepackage{graphicx}%
\usepackage{multirow}%
\usepackage{amsmath,amssymb,amsfonts}%
\usepackage{amsthm}%
\usepackage{mathrsfs}%
\usepackage[title]{appendix}%
\usepackage{xcolor}%
\usepackage{textcomp}%
\usepackage{manyfoot}%
\usepackage{booktabs}%
\usepackage{ulem} 
\usepackage{algorithm}%
\usepackage{algorithmicx}%
\usepackage{algpseudocode}%
\usepackage{listings}%
\usepackage{multirow}
\usepackage{ulem}

\theoremstyle{thmstyleone}%
\theoremstyle{thmstyletwo}%

\theoremstyle{thmstylethree}%

\begin{document}

\title[Article Title]{Scalable Autoregressive Deep Surrogates for Dendritic Microstructure Dynamics}


\author*[1]{\fnm{Kaihua} \sur{Ji}}\email{ji5@llnl.gov}
\equalcont{These authors contributed equally to this work.}

\author[1]{\fnm{Luning} \sur{Sun}}
\equalcont{These authors contributed equally to this work.}

\author[1]{\fnm{Shusen} \sur{Liu}}

\author[1]{\fnm{Fei} \sur{Zhou}}

\author[1]{\fnm{Tae Wook} \sur{Heo}}

\affil[1]{%
  \orgname{Lawrence Livermore National Laboratory}, \orgaddress{%
    \city{Livermore}, \state{CA}, \postcode{94550}, \country{United States of America}%
  }%
}


\abstract{Microstructural pattern formation, such as dendrite growth, occurs widely in materials and energy systems, significantly influencing material properties and functional performance. While the phase-field method has emerged as a powerful computational tool for modeling microstructure dynamics, its high computational cost limits its integration into practical materials design workflows. Here, we introduce a machine-learning framework using autoregressive deep surrogates trained on short trajectories from quantitative phase-field simulations of alloy solidification in limited spatial domains. Once trained, these surrogates accurately predict dendritic evolution at scalable length and time scales, achieving a speed-up of more than two orders of magnitude. Demonstrations in isothermal growth and in directional solidification of a dilute Al-Cu alloy validate their ability to predict microstructure evolution. Quantitative comparisons with phase-field benchmarks further show excellent agreement in the tip-selection constant, morphological symmetry, and primary spacing evolution.}




\maketitle


\section{Introduction}\label{sec1}

Microstructural pattern formation is a ubiquitous phenomenon observed across a wide range of materials and energy systems, with examples including cellular and dendritic solidification during metal casting or additive manufacturing \cite{kurz1989fundamentals,dantzig_solidification_2016,boettinger_phase-field_2002}, phase separation in electrochemical and metallic systems \cite{geslin2015topology,ramana2009study}, and lithium dendrite growth in rechargeable batteries \cite{bai2016transition,chen2015modulation,zhang2023quantitative}. Understanding and accurately predicting microstructural evolution during materials processing and in energy applications is crucial, as it directly informs processing-structure relationships that ultimately determine material properties and functional performance. Despite substantial advances in computational materials science, modeling these microstructural patterns with their coupled multiscale physics and complex morphologies remains a significant challenge.

Among various microstructural patterns, dendritic microstructures that form during alloy solidification processes are particularly important due to their frequent occurrence in alloys with industrial applications and their significant influence on material properties \cite{kurz1989fundamentals,dantzig_solidification_2016,boettinger_phase-field_2002}. The dynamic formation and evolution of dendrites at the solid-liquid interface during solidification are influenced by both intrinsic material properties—including phase equilibria, solute diffusion, and interfacial anisotropies in free energy and kinetics—and extrinsic processing conditions such as growth velocity, temperature gradients, and alloy composition. Capturing the interfacial dynamics across a range of growth conditions is therefore critical for accurately predicting alloy microstructure evolution under both near-equilibrium and far-from-equilibrium regimes at the solid-liquid interface. This is relevant to industrial manufacturing processes ranging from conventional casting, where the solid-liquid interface is near equilibrium \cite{losert1998evolution,Karma2001,echebarria_quantitative_2004}, to advanced additive manufacturing, where it is far from chemical equilibrium \cite{tourret2023morphological,konig2023solidification,ji_microstructural_2023,ji2024microstructure,ji2025phase}.

The phase-field (PF) method has emerged as a powerful computational approach for simulating interfacial problems across various materials and energy systems \cite{boettinger_phase-field_2002,chen2002phase,karma_atomistic_2016,tourret_phase-field_2022}. By employing one or more order parameters to represent different phases and treating interfaces as diffuse transition regions characterized by continuous scalar fields, the PF method avoids explicit interface tracking, making it particularly suitable for modeling interfacial phenomena with complex morphologies. The PF method also facilitates dynamic coupling among multiscale physics, including capillarity, interface kinetics, and transport phenomena. Over the past two decades, quantitative PF simulations of alloy solidification have provided valuable insights into pattern-forming instabilities \cite{tourret2015oscillatory,song_thermal-field_2018}, grain competition \cite{tourret2015growth,Mota2021,song_cell_2023,dorari2022growth}, and fundamental microstructural pattern formation under both near-equilibrium~\cite{Karma2001,echebarria_quantitative_2004} and far-from-equilibrium conditions~\cite{ji_microstructural_2023,ji2024microstructure,ji2025phase}. 
The numerical stability and stiffness of the nonlinear partial differential equations (PDEs) in the PF model typically require small time steps for accurate solutions using various spatial discretization methods (e.g., finite differences, finite elements, and adaptive meshes). In addition, the nanometer-scale width of the solid-liquid interface demands very fine spatial resolution.
Despite significant advances—such as the anti-trapping current~\cite{Karma2001,echebarria_quantitative_2004} and enhanced solute diffusion techniques~\cite{ji_microstructural_2023}—that enable upscaling of the diffuse interface thickness while preserving quantitative accuracy, bridging interfacial to microstructural scales in PF simulations remains computationally expensive. 
This computational cost is particularly limiting for industrially relevant problems that require simulating microstructural evolution over large spatial domains and extended temporal horizons, as seen in casting, metal additive manufacturing, and related alloy solidification processes. Furthermore, the computational intensity often limits the incorporation of PF modeling within Integrated Computational Materials Engineering (ICME) workflows \cite{zhao2023understanding}, which seek rapid turnaround times for material design and optimization tasks \cite{rickman2019materials}.

Recent developments in machine learning (ML), particularly deep learning architectures trained on high-fidelity simulation data for dynamical systems, offer a promising pathway to overcome these limitations. Deep-learning surrogates for PDE-governed systems are commonly categorized as data-driven when data is abundant or physics-informed when data is limited. For instance, the physics-informed neural network (PINN) framework embeds governing equations into the loss function \cite{raissi2019physics}, and has been applied to cardiovascular flows \cite{sun2020surrogate}, turbulence \cite{wang2025simulating,wang2024respecting}, and brittle fracture \cite{goswami2020transfer}. In parallel, operator learning aims to learn mappings between function spaces, improving portability across discretizations and parameters \cite{li2020fourier,lu2021learning}. When abundant datasets are available, data-driven surrogates have achieved strong performance on spatiotemporal prediction using transformers \cite{han2022predicting}, implicit neural representations for high-dimensional reduction \cite{pan2023neural}, and graph-based models for unstructured meshes \cite{pfaff2020learning, bertin2023accelerating, bertin2024learning}. Generative models further capture stochastic behavior and have shown potential in super-resolution, reconstruction, and control \cite{sun2023unifying,gao2024bayesian,du2024conditional,wei2024diffphycon}. Additionally, Large Language Models (LLMs) and Foundation Models (FMs) have shown special capability for complex dynamical systems, such as weather forecasting \cite{bodnar2025foundation,lam2023learning,bi2023accurate}.

Despite these recent advances, there are still considerable obstacles to the large-scale adoption of ML methods for microstructure simulations. For example, PINNs often exhibit slow convergence for time-dependent dynamics and typically require retraining when the discretization or parametrization of the underlying PDE changes. In addition, the operator-based learning usually defines the dataset on a fixed spatial domain. Therefore, it is not a trivial problem for the operator learning to do the spatial extrapolation with zero-shot. Moreover, U-Net-based architectures involve dimensionality reduction during the forward pass. As a result, they can only be combined with domain-decomposition-related techniques for zero-shot extrapolation~\cite{jagtap2020extended,li2019d3m}, but such methods tend to introduce artifacts around the interfaces between sub-domains. In contrast, foundation-model approaches require very large and diverse datasets to generalize reliably, which poses a challenge when high-fidelity data are computationally expensive to generate and experimental labels are scarce.
In the context of microstructure prediction surrogates trained on PF simulations specifically, some require a sequence of consecutive PF frames as input to maintain temporal coherence during prediction~\cite{yang2021self,tseng2023deep}; some rely partially on a physics solver during inference~\cite{montes2021accelerating}; and achieving reliable extrapolation to larger spatial and temporal scales or to unseen thermodynamic and processing conditions remains challenging~\cite{montes2021accelerating, bonneville2025accelerating, fan2024accelerate}.

These constraints are particularly critical because spatial and temporal extrapolation is essential for achieving practical acceleration in microstructure prediction. A typical workflow consists of three coupled stages: data generation (PF simulations), surrogate training, and surrogate prediction. Practical end-to-end speedup requires performance gains combining all three stages.
However, since most surrogate frameworks are trained and evaluated on fixed spatial domains, with limited assessment of spatial generalization, the data generation stage can remain a major bottleneck even when surrogate predictions are computationally efficient. As a result, the realized end-to-end speedup may be limited in practice.

\begin{figure}[htbp]
  \centering
  \includegraphics[width=\textwidth]{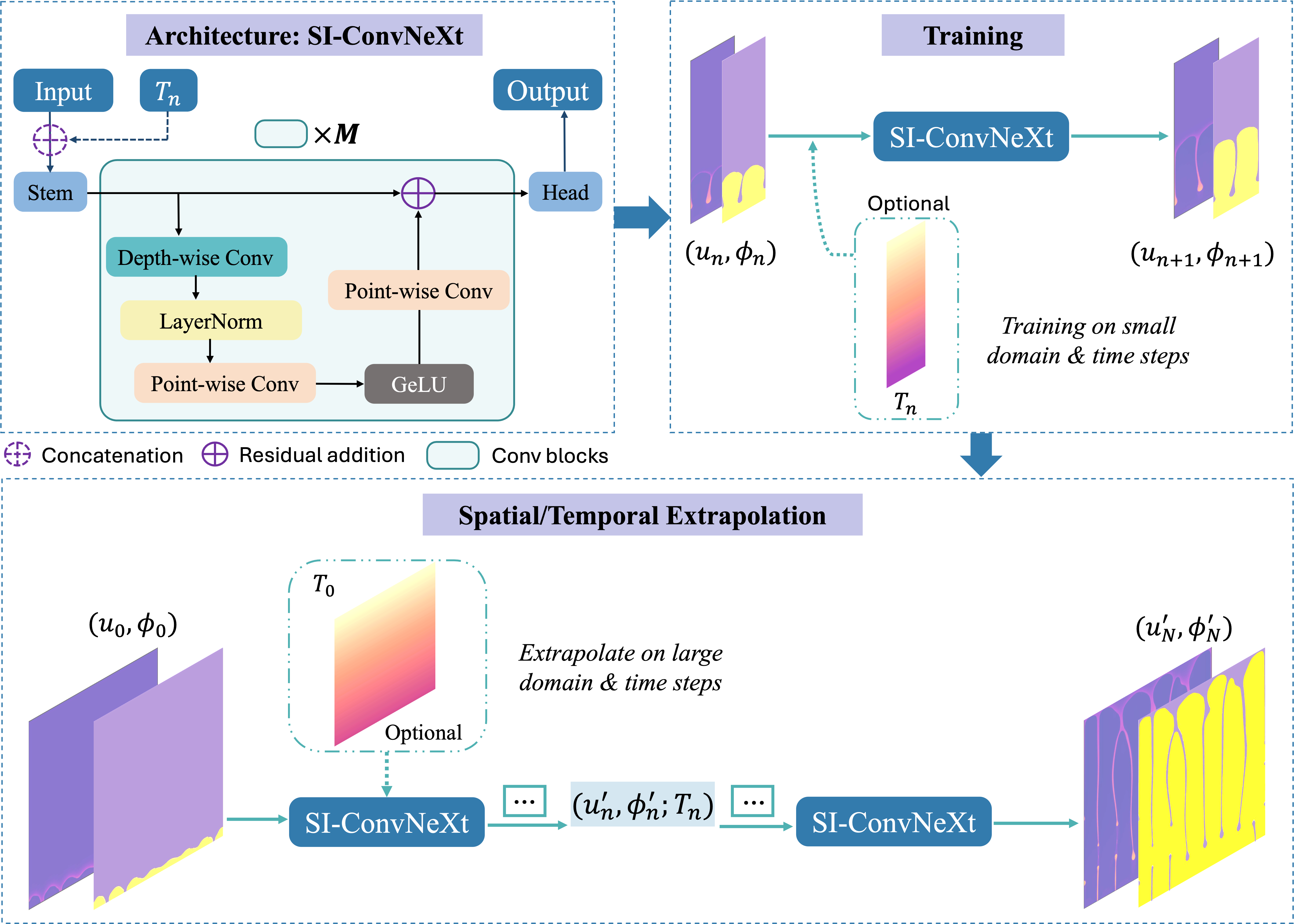}
  \caption{
  The workflow for using autoregressive deep surrogate (ADS) in microstructure prediction. It involves training the Scale-Invariant ConvNeXt (SI-ConvNeXt) model on short microstructural trajectories obtained from quantitative phase-field (PF) simulations conducted over limited spatial and temporal scales. Once trained, the surrogate model can be extrapolated to predict microstructure evolution at larger spatial and temporal scales.
  }
  \label{fig:diagram}
\end{figure}

In this paper, we introduce a ML framework based on temporally autoregressive deep-surrogate (ADS) models, where neural networks learn a one-step time-advance mapping and are iteratively rolled out to accurately predict microstructure evolution. This approach circumvents the aforementioned limitations and enables scalable, efficient microstructure prediction. We build our model upon ConvNeXt~\cite{liu2022convnet} inspired by the design of vision transformers~\cite{dosovitskiy2020image}. 
To improve spatial extrapolation ability, we modify the architecture to obtain a Scale-Invariant ConvNeXt (SI-ConvNeXt) by maintaining resolution without spatial dimension reduction and integrate this modified architecture into a workflow, as shown in Fig.~\ref{fig:diagram}. Simulation results demonstrate that this design is critical for scaling and for accurately predicting translationally invariant microstructure evolution in isothermal solidification. While dimension-reduction techniques such as U-Net have become mainstream, the associated compression and loss of spatial resolution are detrimental to dendrite-growth predictions that require zero-shot spatial extrapolation; we therefore adopt a resolution-preserving design.
Our model is also flexible in incorporating conditional fields, such as temperature, to inform alloy solidification under a temperature gradient, where the microstructure evolution becomes non-translationally invariant. This conditioning introduces physical variables into the input without adding PDE residuals or hard constraints. As such, the approach remains data-driven while being trained on physics-based PF simulations. Unlike models that require sequences of input frames for spatial and temporal extrapolation, our one-step autoregressive surrogate predicts microstructure evolution based solely on the initial condition. These surrogates are trained on microstructural trajectories obtained from quantitative PF simulations conducted in spatially limited domains. Once trained, the models are capable of predicting microstructure evolution from a single initial frame, without further involvement of the physical solver, and can extrapolate across both spatial and temporal scales.

We demonstrate this approach using two representative cases: 
(i) isothermal dendrite growth under near-equilibrium conditions, and 
(ii) directional solidification under far-from-equilibrium conditions. Quantitative comparisons between PF benchmarks and ADS predictions reveal excellent agreement in tip-selection constants and morphological symmetry for single dendrite growth under isothermal conditions.
In addition, simulations of multiple dendrites in different domain sizes demonstrate that the extrapolation accuracy is independent of the domain size, allowing extrapolation at experimentally relevant length scales. Temporal extrapolation, however, is more constrained due to error accumulation: small per-step prediction errors accumulate over successive autoregressive steps, gradually degrading accuracy over long prediction horizons. We further analyze the effects of key training hyperparameters, including training dataset size and time intervals between auto-regressive steps, and suggest an optimal training strategy.
For directional solidification, the surrogates accurately capture the evolution of primary spacing, achieving a simulation speed-up exceeding two orders of magnitude. These results demonstrate that the SI-ConvNeXt-based ADS framework not only performs well under translationally invariant dynamics (e.g., isothermal growth) but also generalizes effectively to non-translationally invariant systems (e.g., directional solidification).

The presented workflow provides a viable route toward scalable microstructure prediction, i.e., first have the neural network learn the localized dynamics governed by complex PDEs within limited spatial domains and time steps, for which data can be quickly generated from high-fidelity physical solvers; then, use the ADS to efficiently model the multiscale microstructure evolution at experimentally relevant length and time scales. This paves the way for the integration of computationally expensive PF simulations into ICME workflows and practical alloy design by providing critical processing-structure relationships.
The successful application of this scalable autoregressive surrogate approach in accurately modeling dendritic microstructure evolution in the two representative solidification scenarios presented herein demonstrates its potential applicability to a broad range of microstructural pattern-forming phenomena beyond alloy solidification, including the modeling of phase transformations in a wide range of materials and energy systems.

\section{Results}\label{sec2}

\subsection{Isothermal dendrite growth}

We first consider the case of isothermal dendrite growth. To obtain simulation data for training the ADS, we use a well-established quantitative PF model for solidification of dilute binary alloys~\cite{Karma2001,echebarria_quantitative_2004}.
The model utilizes thin-interface asymptotics \cite{karma_quantitative_1998,echebarria_quantitative_2004} to make the diffuse interface width $W$ much larger than the capillarity length $d_0$ while remaining quantitative.
An anti-trapping current is introduced into the model to compensate for spurious solute trapping due to a large $W$ \cite{Karma2001}. The local equilibrium conditions are restored at the solid-liquid interface, making the PF model applicable to solidification processes such as casting and directional solidification where the interface growth rate typically lies around $100~\mu\mathrm{m/s}$ or below \cite{boettinger2000solidification,li2012dendrite,clarke_microstructure_2017}.

In the model, a variable $\phi$ (the phase field) smoothly interpolates between the solid ($+1$) and liquid ($-1$) phases. To enhance numerical stability at larger grid spacings~\cite{glasner2001nonlinear}, a preconditioned phase field $\psi$ is used in the simulations, which is related to $\phi$ by:
\begin{equation}
\phi(\mathbf{r}, t)=\tanh\left\{{\psi(\mathbf{r}, t)}/{\sqrt{2}}\right\}.
\label{precondition}
\end{equation}
This PF variable $\psi$ is coupled to a dimensionless supersaturation field $U$, which measures deviations from chemical equilibrium at the interface. With known $\psi$, $U$ can be translated to and from the solute concentration field $c$.
Under the isothermal condition, the dynamics for dendritic solidification are translationally invariant; that is, the evolution equations for $\psi$ and $U$ remain consistent across the domain. 
We solve the evolution equations numerically using finite-difference discretizations with explicit Euler time stepping.
Because only neighboring grid points are involved in updating the $\psi$ and $U$ fields without any long-range interactions, convolutional layers with minimal $3 \times 3$ kernels in the SI-ConvNeXt architecture are expected to effectively capture these localized dynamics.

\begin{figure*}[htbp]
  \centering
  \includegraphics[scale=0.6]{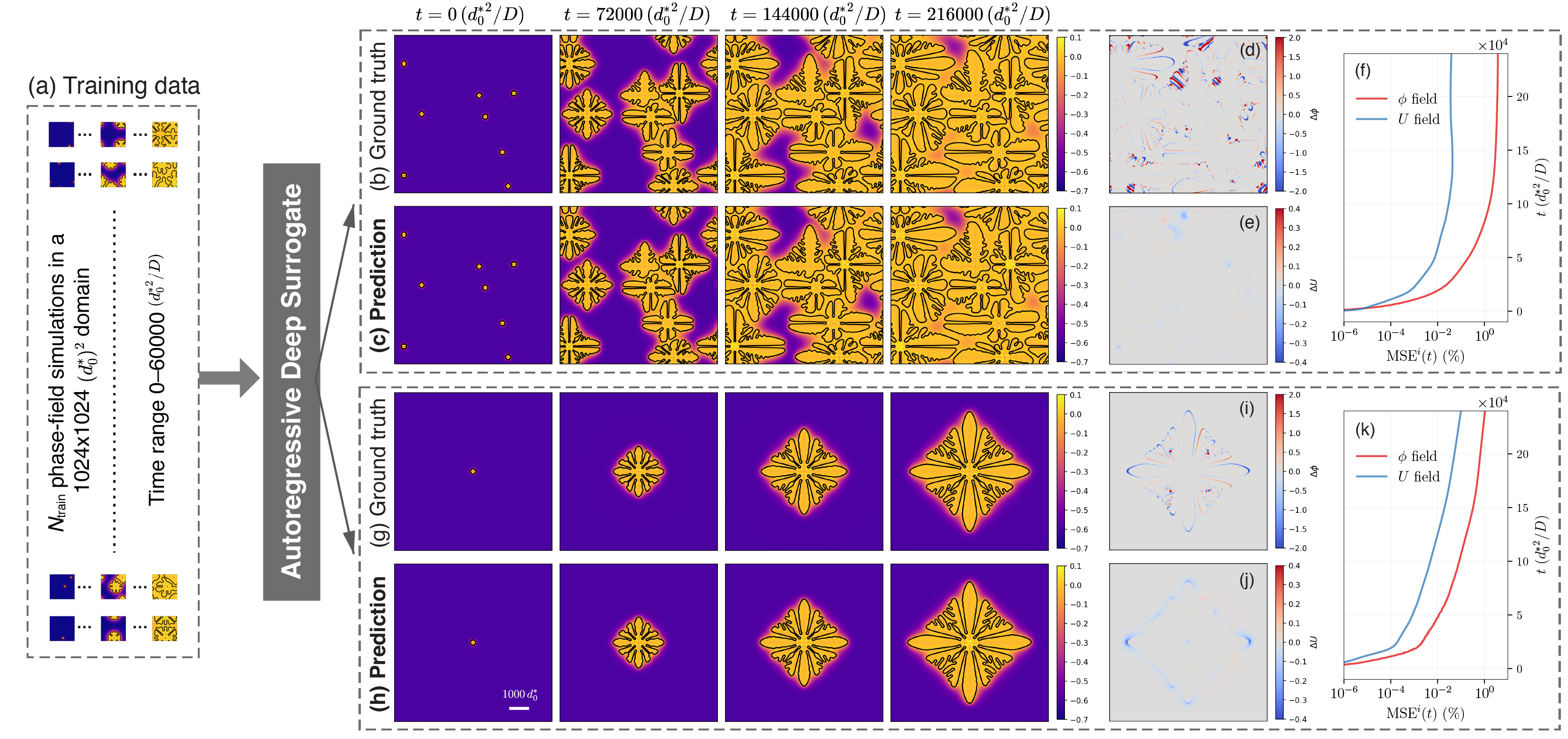}
  \caption{Predicting isothermal dendrite growth using ADS. (a) Training data are generated from $N_{\mathrm{train}} = 20$ PF simulations on a domain of size $1024 \times 1024\, (d_0^*)^2$ over a simulated time of $t_{\mathrm{tot}}^{\mathrm{train}} = 6 \times 10^4\, (d_0^*)^2 / D$. Once trained, the ADS is used to predict microstructure evolution at larger spatial and temporal scales. 
  (b) Ground truth results from PF simulations of multiple dendrite growth. (c) Predicted microstructure using the trained ADS. In both (b) and (c), black contours represent $\phi = 0$, and the colormap denotes the $U$ field.
  (d)-(e) Spatial distributions of prediction errors $\Delta \phi$ and $\Delta U$ at $t = 2.16 \times 10^5\, (d_0^*)^2 / D$. (f) Spatially averaged mean-squared error $\mathrm{MSE}^i(t)$ for both $\phi$ and $U$ as a function of time.
  (g)-(k) Corresponding results for a single dendrite growth predicted using the same trained ADS model, analogous to (b)-(f).
  }
  \label{fig:dendrite}
\end{figure*}

Figure~\ref{fig:dendrite}(a) illustrates the preparation of the training data, in which PF simulations are conducted on relatively small domains of size $1024 \times 1024\,(d_0^*)^2$ (corresponding to $256 \times 256$ grid points), where $d_0^*$ is the capillary length defined at the isothermal temperature. 
We randomly initialize 1–4 square-shaped solid seeds, each with a side length between $6$ and $24\,d_0^*$, and randomly position them within the simulation domain. All seeds have the same crystallographic orientation aligned with vertical and horizontal axes, and the interfacial free-energy anisotropy exhibits four-fold symmetry with $\epsilon_4=0.007$.
The initial $U$ field is set to $U = -\Omega (1 - \phi)/2$, with a supersaturation $\Omega = 0.6$. Periodic boundary conditions are applied on all sides. There are $N_{\mathrm{train}}$ training simulations, each running up to a time of $t_{\mathrm{tot}}^{\mathrm{train}} = 6 \times 10^4\, (d_0^*)^2/D$. Simulation fields are saved every $R_t$ time steps, where the base PF time step is $\Delta t_{\mathrm{PF}} = 2.4\, (d_0^*)^2/D$ and the saving interval corresponds to $\Delta t = 240\, (d_0^*)^2/D$ (i.e., $R_t \equiv \Delta t / \Delta t_{\mathrm{PF}} = 100$).
The final training dataset consists of the trajectories of the $\phi$ and $U$ fields from $N_{\mathrm{train}}$ simulations with different initial conditions, where the simulation output $\psi$ is converted to $\phi$ using Eq.~\eqref{precondition}, such that the phase field varies between $[-1, 1]$. The data are then downsampled to $64 \times 64$ grid points by averaging over $4 \times 4$ blocks of the original grid points.
Exploiting the four-fold symmetry (point group $4m$, encompassing four rotations and four reflections), data augmentation strategies are employed during the training of the ADS based on the SI-ConvNeXt architecture (see Methods for details).

Once trained, the ADS can predict the spatiotemporal evolution of microstructures, generating full trajectories $\{\phi_1, \ldots, \phi_N\}$ and $\{U_1, \ldots, U_N\}$ given only the initial conditions $(\phi_0, U_0)$. At each step, the ADS outputs $(\phi_n, U_n)$ are used as inputs to predict the next state $(\phi_{n+1}, U_{n+1})$, where the time step interval is $\Delta t$.
The trained ADS model is then tested by extrapolating its predictive capabilities to larger spatial domains and longer time horizons. As a first demonstration, eight randomly sized and positioned square seeds are placed within a larger simulation domain of size $6400 \times 6400\, (d_0^*)^2$, with $\Omega = 0.6$. 
As shown in the ground truth (GT) PF simulation results [Fig.~\ref{fig:dendrite}(b)], each seed initially grows along the four crystallographically preferred directions imposed by the four-fold anisotropy in the interfacial free energy. Depending on the initial seed size, dendrite tips evolve into either single-tip morphologies or doublon structures (two closely spaced dendrite tips separated by liquid channels) \cite{utter2005double}. This initial morphological selection influences subsequent competitive dendrite growth.

Remarkably, using only the initial fields at $t = 0$, the ADS accurately predicts microstructure evolution up to $t_{\mathrm{tot}}^{\mathrm{pred}} = 2.4 \times 10^5\, (d_0^*)^2 / D$ (500 autoregressive steps). As shown in Fig.~\ref{fig:dendrite}(c), the predicted (PD) microstructure closely matches the GT results, accurately capturing both dendrite tip morphologies and competitive growth dynamics. The spatial distributions of the prediction errors $\Delta \phi$ and $\Delta U$ at $t = 2.16 \times 10^5\, (d_0^*)^2 / D$ are shown in Fig.~\ref{fig:dendrite}(d)-(e), which reveal only minor discrepancies between the PD and GT results.
To quantify prediction accuracy over the entire prediction horizon, we calculate the point-wise mean-squared error (MSE) normalized by the initial field range. Specifically, for each field $i \in \{\phi, U\}$, we compute the normalized MSE averaged over the entire spatial domain at each stored time step $t_n$, denoted as $\mathrm{MSE}^i(t)$ (see Methods for details), and plot it in Fig.~\ref{fig:dendrite}(f). The values of $\mathrm{MSE}^i(t)$ for both fields gradually increase and eventually saturate as dendritic growth fills the simulation domain. Overall, the $\phi$ field exhibits slightly higher error than the $U$ field, but $\mathrm{MSE}^\phi(t)$ for both fields remains around 1\% throughout the simulation.

\begin{figure}[htbp]
  \centering
  \includegraphics[scale=0.48]{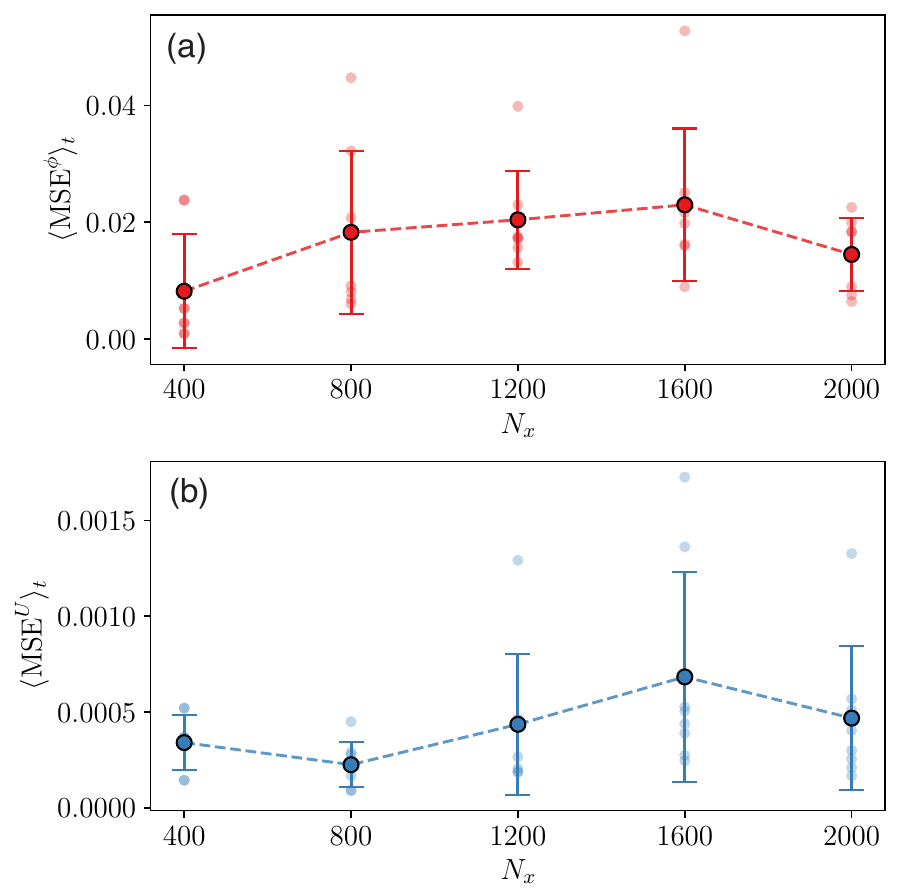}
  \caption{The normalized mean-squared error: (a) $\bigl\langle \mathrm{MSE}^\phi \bigr\rangle_t$ and (b) $\bigl\langle \mathrm{MSE}^U \bigr\rangle_t$ between the ground truth and predicted results for the $\phi$ and $U$ fields, respectively, plotted as a function of the extrapolation domain size $N_x$ (i.e., the number of grid points along the $x$-direction in a square simulation domain). Five surrogates are trained on the same dataset in a small domain size shown in Fig.~\ref{fig:dendrite}(a). Each data point is the average of five ADS simulations using the five trained surrogates, and the error bars represent the standard deviation.
  }
  \label{fig:dendrite_extrapolate}
\end{figure}

To assess the scalability of the trained ADS model, we further performed extrapolated modeling across different simulation domain sizes, varying the side length $L$ from $1600$ to $8000\, d_0^*$ (corresponding to $400$ to $2000$ grid points in PF simulations). We use a single value, the space- and time-averaged mean-squared error $\bigl\langle \mathrm{MSE}^i \bigr\rangle_t$, to quantify the overall ADS prediction accuracy (see Methods for details).
As shown in Fig.~\ref{fig:dendrite_extrapolate}, these results demonstrate that both $\bigl\langle \mathrm{MSE}^\phi \bigr\rangle_t$ and $\bigl\langle \mathrm{MSE}^U \bigr\rangle_t$ remain low (approximately 2\% for $\phi$ and 0.05\% for $U$) and nearly constant across the tested domain sizes, with only a slight increase at large $N_x$. This indicates robust spatial scalability within the explored range.
However, extrapolation over extended temporal horizons presents inherent limitations. Due to the autoregressive nature of the prediction process, errors accumulate over successive prediction steps and gradually degrade accuracy. Although we introduce a small amount of Gaussian noise to the input during training to improve the robustness of predictions, the accumulated errors can eventually lead to noticeable deviations in the predicted dendritic morphology compared to GT simulations. As shown in Fig.~\ref{fig:dendrite}(f), such errors increase at sufficiently long prediction times but remain small within the temporal extrapolation achieved in the PD simulation, where $t_{\mathrm{tot}}^{\mathrm{pred}} / t_{\mathrm{tot}}^{\mathrm{train}} = 4$.
Strategies to mitigate this temporal error accumulation, such as hybrid approaches combining surrogate predictions with periodic PF solver corrections~\cite{um2020solver,fan2025diff}, could further extend the temporal reliability and enhance predictive capability over long simulation times. However, such techniques may also compromise the performance efficiency of the ADS, which warrants future investigation.

\begin{figure*}[htbp]
  \centering
  \includegraphics[scale=0.75]{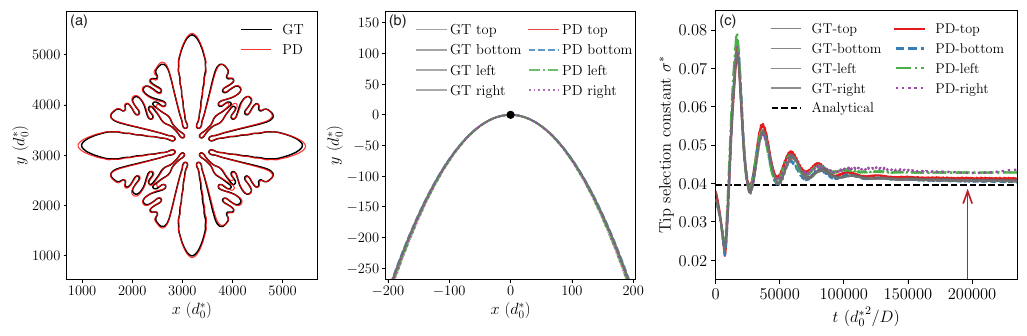}
  \caption{(a) Comparison of the $\phi = 0$ contour of a single dendrite at $t = 1.96 \times 10^5\, (d_0^*)^2 / D$ from the GT and PD simulations. (b) Enlarged views of the dendrite tip regions from all dendrites in the GT and PD simulations, where the tip locations are aligned through rotation and translation to a common reference point. (c) Temporal evolution of the tip selection constant $\sigma^*$ for all dendrites in the GT and PD simulations. The black dashed line represents the analytical value $\sigma^* = 0.0396$, and the arrow indicates the time corresponding to the dendrite shown in panels (a) and (b).}
  \label{fig:dendrite_quantify}
\end{figure*}

We further evaluate the performance of the trained ADS by examining the growth of a single dendrite in a larger simulation domain. Figure~\ref{fig:dendrite}(g) shows the GT simulation result, where a single circular solid seed with an initial radius of $40\,d_0^*$ is placed at the domain center under an initial supersaturation $\Omega = 0.6$. Due to the imposed four-fold interfacial anisotropy, the dendrite initially develops four primary arms aligned with the crystallographically preferred growth directions, and subsequently forms secondary arms and complex sidebranching structures.
As shown in Fig.~\ref{fig:dendrite}(h), the ADS trained on the same dataset used for predicting multiple dendrite growth in Fig.~\ref{fig:dendrite}(c) also accurately predicts the evolution of a single large dendrite. The spatial distributions of the prediction errors $\Delta \phi$ and $\Delta U$ at $t = 2.16 \times 10^5\, (d_0^*)^2 / D$ are shown in Fig.~\ref{fig:dendrite}(i)-(j). These errors remain localized primarily near the dendrite tip regions and along secondary branching regions, with negligible discrepancies observed in the bulk solid and liquid phases.
Furthermore, the temporal evolution of $\mathrm{MSE}^i(t)$ in Fig.~\ref{fig:dendrite}(k) demonstrates excellent prediction accuracy, with normalized errors maintained around $1\%$ for the $\phi$ field and below $0.1\%$ for the $U$ field.
A direct comparison of dendrite shapes between the GT and PD simulations is provided in Fig.~\ref{fig:dendrite_quantify}(a), where contours corresponding to the solid-liquid interface ($\phi = 0$) from both simulations are superimposed. Minor deviations in the dendrite contours are observed only in the sidebranching regions and as small asymmetries in the left and right primary dendrite arms. A detailed view of the superimposed dendrite tips of the four arms from both GT and PD simulations [Fig.~\ref{fig:dendrite_quantify}(b)] further highlights the excellent agreement in dendrite tip morphology.

To quantify dendrite tip dynamics, we compute the tip selection constant $\sigma^*$, defined as
\begin{equation}
\sigma^* = \frac{2 D d_0^*}{\rho^2 V}, \label{sigma}
\end{equation}
where $\rho$ is the dendrite tip radius and $V$ is the tip growth velocity. At low undercooling, $\sigma^*$ converges to a constant value that depends solely on the strength of interfacial anisotropy~\cite{barbieri_predictions_1989,Karma2000}.
To measure $\sigma^*$, we track the position of the most advanced solid-liquid interface along each of the four crystallographically preferred growth directions and extract the tip locations $(x_{\mathrm{tip}}, y_{\mathrm{tip}})$ to determine $V$. The tip radius $\rho$ is obtained by fitting the local interface shape to a parabola, $y=y_{\mathrm{tip}}-(x-x_{\mathrm{tip}})^2/(2\rho)$, following the procedure described in Refs.~\cite{clarke_microstructure_2017,ji_isotropic_2022}. With the measured values of $\rho$ and $V$, we evaluate $\sigma^*$ using Eq.~\eqref{sigma}.
The temporal evolution of $\sigma^*$ for the four dendrite arms is shown in Fig.~\ref{fig:dendrite_quantify}(c), comparing results from both GT and PD simulations with the analytical value $\sigma^* = 0.0396$ calculated from solvability theory \cite{barbieri_predictions_1989} for $\epsilon_4=0.007$. In the GT simulation, all four arms exhibit identical transient behavior and converge to the analytical $\sigma^*$ value. The ADS predictions closely follow these transients and reach steady-state $\sigma^*$ values within approximately $5\%$ of the analytical solution.
Notably, it takes approximately $10^5\, (d_0^*)^2 / D$ for $\sigma^*$ to reach its steady-state value, whereas the training data span only a limited duration of $6 \times 10^4\, (d_0^*)^2 / D$ and a confined spatial domain where dendrite tips do not reach steady state. Nevertheless, the ADS successfully predicts the correct steady-state $\sigma^*$ value, despite this quantitative feature not being present in the training data. This further demonstrates that the ADS has effectively learned the local interfacial dynamics governed by the PF model PDEs, even without explicit knowledge of the underlying equations.

\begin{figure*}[htbp]
  \centering
  \includegraphics[width=\textwidth]{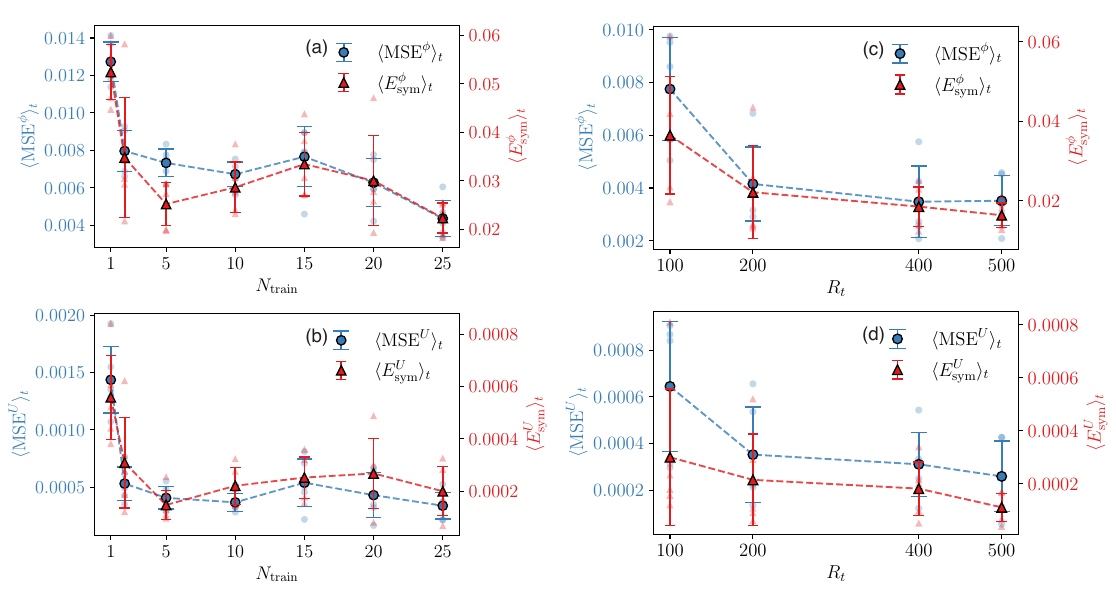}
  \caption{Time-averaged mean-squared errors $\langle \mathrm{MSE}^{i} \rangle_t$ and time-averaged symmetry errors $\langle E_{\mathrm{sym}}^{i} \rangle_t$ as functions of the number of training simulations ($N_{\mathrm{train}}$) and the ratio $R_t \equiv \Delta t / \Delta t_{\mathrm{PF}}$, where $\Delta t$ is the ADS prediction time step and $\Delta t_{\mathrm{PF}}$ is the time step of the PF finite-difference solver. Five surrogates are trained on the same dataset in a small domain size shown in Fig.~\ref{fig:dendrite}(a). Each data point is the average of five ADS simulations using the five trained surrogates, and the error bars represent the standard deviation.}
  \label{fig:dendrite_MSE}
\end{figure*}

We next investigate the influence of two important hyperparameters—the number of training simulations $N_{\mathrm{train}}$ and the prediction time interval $\Delta t$—on the prediction accuracy and symmetry of a single dendrite. Quantitative assessments are performed using the space- and time-averaged mean squared error, $\langle \mathrm{MSE}^{i} \rangle_t$, and the time-averaged symmetry error, $\langle E_{\mathrm{sym}}^{i} \rangle_t$ (see Methods for details). 
For a given $N_{\mathrm{train}}$, we train five surrogate models with different random seeds and use them to predict the same single-dendrite growth scenario shown in Fig.~\ref{fig:dendrite}(g). As illustrated in Fig.~\ref{fig:dendrite_MSE}(a)-(b), both $\langle \mathrm{MSE}^{i} \rangle_t$ and $\langle E_{\mathrm{sym}}^{i} \rangle_t$ are evaluated for $i \in \{\phi, U\}$. When $N_{\mathrm{train}} = 1$, both error metrics are high, but they decrease rapidly with increasing $N_{\mathrm{train}}$ and saturate around $N_{\mathrm{train}} = 5$. This indicates that only a few short PF simulations in small domains are sufficient for the ADS to learn the essential interfacial dynamics.
In addition, we assess the impact of the ADS time step by considering the ratio $R_t \equiv \Delta t / \Delta t_{\mathrm{PF}}$ between the time steps used in ADS predictions and in the finite-difference solver for PF. For each $R_t$, five surrogate models are trained with different random seeds and used to predict the same dendrite growth. Unlike finite-difference solvers, which are constrained by numerical stability, ADS predictions do not have such limitations. However, if $R_t$ exceeds a critical value $R_t^c \approx 550$, the ADS fails to capture the PF dynamics due to excessive temporal decorrelation. Conversely, for $R_t < R_t^c$, as shown in Fig.~\ref{fig:dendrite_MSE}(c)-(d), prediction accuracy improves as $\Delta t$ increases, since fewer autoregressive steps reduce cumulative errors.
In summary, to optimize accuracy and computational efficiency, $N_{\mathrm{train}}$ should be chosen near the saturation point (here, $N_{\mathrm{train}} \approx 5$), and $R_t$ should be selected close to but below the critical threshold $R_t^c$.

\subsection{Dendrite growth within a temperature gradient}

\begin{figure*}[htbp]
  \centering
  \includegraphics[scale=0.9]{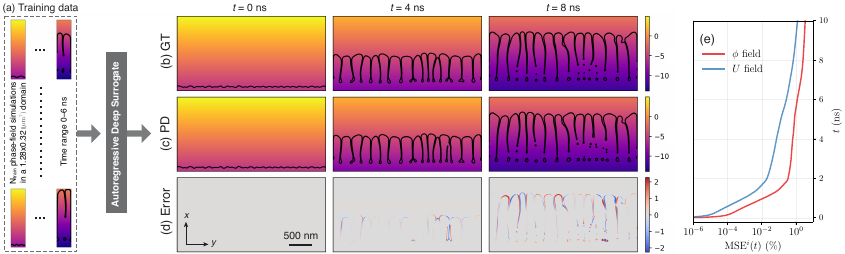}
  \caption{
  Predicting dendrite growth within a temperature gradient under additive manufacturing conditions using ADS. (a) the training data is prepared from $N_{\mathrm{train}}=24$ phase-field simulations in a domain of size $L_x \times L_y = 1.28 \times 0.32\,\mathrm{\mu m^2}$ over a simulated time 0-6 ns. Compare the (b) GT and (c) PD simulations in a domain of size $L_x \times L_y = 5.12 \times 0.64\,\mathrm{\mu m^2}$ over 0-10 ns, where $G=52.80$ $\mathrm{K/\mu m}$ and the initial $T_{\mathrm{interf}}=-5.02$. The black contours represent $\phi=0$, and the colormap represents the temperature field (in normalized unit).
  (d) Spatial distribution of errors in terms of $\Delta \phi$ at different time. (e) Spatially averaged mean-squared error $\mathrm{MSE}^i(t)$ for both $\phi$ and $c$ as a function of time.}
  \label{fig:AM}
\end{figure*}

Microstructure evolution typically occurs under the influence of external fields. A prominent example is directional solidification \cite{trivedi1980theory,losert1998evolution,song_cell_2023}, which has been widely utilized to produce columnar and single-crystal alloys. In this process, the microstructure evolves under a controlled temperature gradient and interface growth rate. Solidification with moving isotherms is also characteristic of rapid solidification processes such as laser powder bed fusion \cite{tourret2023morphological,konig2023solidification}, where a high-energy laser locally melts and resolidifies the metal powder, generating steep thermal gradients and rapidly moving isotherms.
Here, we consider alloy solidification under a temperature gradient using the frozen temperature approximation (FTA), which assumes that thermal diffusion is much faster than solutal diffusion and can thus be considered effectively ``frozen.'' Under the FTA, the temperature field directed along the interface growth direction ($+x$) is described as $T = T_0 + G\left(x - V_{\mathrm{iso}} t\right)$,
where $T_0$ is a reference temperature, chosen as the equilibrium liquidus temperature. 
This expression is derived in a local frame of reference (the ``material frame''), which remains stationary with respect to the solidifying material. Within this frame, isotherms move at a constant speed $V_{\mathrm{iso}}$ in the $+x$ direction. Given prescribed values of $G$ and $V_{\mathrm{iso}}$, the spatiotemporal evolution of the temperature field is predefined.

Under a uniform temperature field, the ADS can predict the next frame of isothermal dendrite growth based solely on the current frame consisting of two channels, $(\phi_n, c_n)$, where $c$ denotes the concentration field (i.e., the molar fraction of a component in a binary alloy).
However, for directional solidification, given the current frame $(\phi_n, c_n)$, it is insufficient for the ADS to predict the next time step without being informed of the thermal condition. Accordingly, we incorporate a third channel for the temperature field into each input frame, such that the input to the ADS at time step $n$ becomes $(\phi_n, c_n; T_n)$ for predicting the next step $(\phi_{n+1}, c_{n+1}; T_{n+1}^\prime)$. Since the temperature field is predefined, it is excluded from the loss calculation during training. During prediction, the temperature component $T_{n+1}^\prime$ output by the ADS is discarded, and the predefined $T_{n+1}$ is concatenated with the predicted $\phi_{n+1}$ and $c_{n+1}$ to form the input for the next autoregressive step, i.e., $(\phi_{n+1}, c_{n+1}; T_{n+1})$.
A more general approach is to treat the temperature field in the same manner as the $\phi$ and $c$ fields, i.e., including it in the loss function and propagating it through autoregressive steps. This strategy would enable the modeling of alloy solidification coupled with latent heat diffusion and dynamic thermal field evolution \cite{song_thermal-field_2018,ji2025phase}, which, however, is beyond the scope of this paper and will be explored in future work.

In this case study, we investigate dendritic microstructure evolution under far-from-equilibrium conditions, which are relevant to rapid solidification processes such as laser powder bed fusion and laser cladding. Modeling microstructure evolution during rapid solidification is particularly important for understanding the processing-structure relationships in advanced manufacturing.
To generate training date, we consider the rapid solidification of a dilute Al-3\% Cu alloy and use the state-of-the-art PF model for far-from-equilibrium alloy solidification \cite{ji_microstructural_2023,ji2025phase}. This model quantitatively incorporates key nonequilibrium phenomena at the solid-liquid interface, including solute trapping and solute drag, and have successfully predicted the microstructure pattern transition near the analytically predicted stability limits at extremely high velocities \cite{ji_microstructural_2023,ji2024microstructure}. 
In this formulation, the evolution equations for the phase field $\phi$, which varies smoothly from $+1$ in the solid to $-1$ in the liquid, is coupled to a concentration field $c$ that is scaled by the nominal solute concentration $c_{\infty}$. We consider a physically realistic solid-liquid interface thickness ($W=W_0$), typically on the order of nanometer, which offers the highest fidelity by accurately representing interfacial phenomena without introducing spurious interface effects or requiring additional corrective terms~\cite{ji_microstructural_2023,ji2025phase}. However, bridging the interfacial to microstructural length and time scales using this PF formulation is computationally demanding, making this problem well-suited for acceleration via surrogate modeling.

As illustrated in Fig.~\ref{fig:AM}(a), the training dataset is generated using 2D PF simulations conducted in a domain of size $L_x \times L_y = 1.28 \times 0.32\,\mathrm{\mu m^2}$ over a time horizon from 0 to 6~ns. The initial conditions are prepared as follows.
First, we compute approximate 1D steady-state solutions of the PF model at a fixed isotherm velocity $V_{\mathrm{iso}} = 0.12$~m/s, following the procedure described in Ref.~\cite{ji2025phase}. These steady-state solutions correspond to an interface temperature $\widetilde{T}_{\mathrm{interf}}$, where $\widetilde{T} = (T - T_M)/(m_e c_{\infty})$ is the scaled temperature (hereafter, we denote $\widetilde{T}$ simply as $T$). This provides equilibrium profiles of the phase field and concentration field across the solid–liquid interface, which are then extended into 2D.
Then, we perturb the planar solid–liquid interface in 2D by imposing a sinusoidal wave with a single random wavelength ranging from 0.08 to 0.32~$\mu$m. This is done by laterally shifting the entire 1D profiles in the $x$-direction. Additionally, the temperature $T_{\mathrm{interf}}$ at the most advanced interface in the $x$-direction is varied between $-6$ and $-5$ (in normalized units), while the temperature gradient $G$ is sampled uniformly in the range 10–100~K/$\mu$m.
In total, the training dataset comprises $N_{\mathrm{train}} = 24$ independent simulations, encompassing a diverse set of initial perturbations and thermal conditions.

\begin{figure}[htbp]
  \centering
  \includegraphics[scale=1]{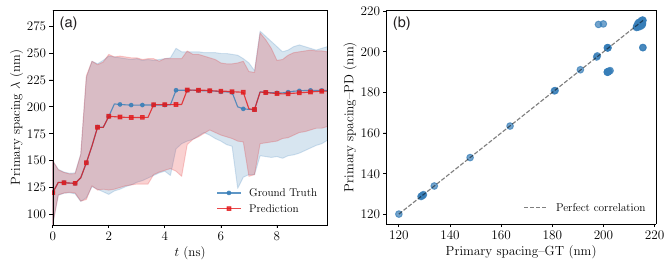}
  \caption{(a) Evolution of the primary spacing $\lambda$ in GT and PD simulations corresponding to Fig.~\ref{fig:AM}(b)-(c), where the dots and lines represents the average $\lambda$ and the colored regions represent the range of $\lambda$. (b) The correlation of PD and GT spacings, where each dot represents the average $\lambda$ at a certain time.
  }
  \label{fig:AM_spacing}
\end{figure}

We perform PD simulations using the trained ADS in a larger simulation domain ($L_x \times L_y = 5.12 \times 0.64\,\mathrm{\mu m^2}$) over extended time horizons of up to 8 ns. The domain is extended in both $x$ and $y$ directions, with a primary extension in the transverse ($y$) direction to examine the dynamics of primary spacing, which will be discussed shortly.
The initial solid-liquid interfaces in the PD simulations are perturbed by a composite sinusoidal wave (in contrast to the single-wavelength perturbation used in the training datasets):
\begin{equation}
\Delta(y) \;=\; \frac{A_0}{2} \sum_{k=1}^{5} \alpha_k \,
\sin\!\biggl(\frac{2\pi\,y}{\lambda_k}\biggr),
\qquad 0 < y < L_y,
\label{eq:interface_perturbation}
\end{equation}
where $\Delta(y)$ is the displacement of the interface position in the $x$-direction, each wavelength $\lambda_k$ and weight $\alpha_k$ are independently and uniformly sampled from the intervals $\lambda_k \in [0.04,\, 0.32]$ $\mu\mathrm{m}$ and $\alpha_k \in [0.2,\,1.0]$, respectively. The global amplitude $A_0$ is drawn from the range $[8,\,16]$ $\mathrm{nm}$. 
Solidification processing conditions are imposed via the predefined temperature field, by fixing the isotherm velocity $V_{\mathrm{iso}}=0.12$ m/s and sampling the thermal gradient $G \in [10,\,100]$ $\mathrm{K/\mu m}$ along with an initial interface temperature $T_{\mathrm{interf}} \in [-6,\,-5]$.
Figure~\ref{fig:AM}(c) shows the ADS prediction in comparison with the GT simulation result shown in Fig.~\ref{fig:AM}(b). In both simulations, the interface evolves into dendritic structures. Visually, the PD simulation accurately captures the evolution of dendrite morphology, reproducing dendrite elimination, branching events, and other transient changes observed in the GT simulation.
The spatial distribution of the prediction error $\Delta \phi$ at the final time step is shown in Fig.~\ref{fig:AM}(d), revealing minor discrepancies localized near the interface. Additionally, the normalized mean-squared errors $\mathrm{MSE}^i(t)$ for both $\phi$ and $c$ as functions of time are shown in Fig.~\ref{fig:AM}(e), which gradually increase over time, reaching approximately $1\%$ at later stages. This demonstrates that the ADS maintains high accuracy throughout prolonged predictions, despite the complex dynamics underlying microstructure evolution.

We further quantify microstructure evolution by measuring the primary dendrite spacing $\lambda$ as a function of simulation time. To obtain accurate spacing measurements, we first identify the solidification front at each time frame by scanning each vertical ($x$) column of the 2D simulation grid and performing linear interpolation at the interface position where $\phi = 0$. Spacings are measured along a fixed horizontal line located $64$\,nm below the mean interface height. Solid regions (dendrites) are identified as contiguous segments with $\phi > 0$, and dendrite centers are computed as the midpoints between segment endpoints. Local spacings are then determined as the distances between neighboring dendrite centers, with each dendrite assigned the average of the spacings to its immediate left and right neighbors.
In early growth stages, characterized by shallow grooves and poorly defined dendrite columns, a secondary peak-to-peak spacing method is employed, where local maxima along the interface are identified and spacings are calculated from their positions. By combining these two methods, the full temporal evolution of the primary spacing is accurately measured.
The evolution of the mean primary spacing measured from both GT and ADS predictions is shown in Fig.~\ref{fig:AM_spacing}(a). Excellent agreement is observed during the early stages of the simulation, with the ADS accurately capturing dendrite branching and elimination events that drive spacing adjustments during transient growth. Although minor discrepancies in event timing emerge in later stages, the overall morphological evolution remains well captured.
To further assess prediction accuracy, we perform a correlation analysis between the GT and PD spacings. For each time step, pairs of spacings $(S_{\mathrm{GT}}, S_{\mathrm{PD}})$ are extracted, and the Pearson correlation coefficient $R$ is computed. Squaring this value gives the coefficient of determination, $R^2$, which quantifies the proportion of variance in the PD spacings that is explained by the GT spacing. Figure~\ref{fig:AM_spacing}(b) presents the correlation results for the primary spacings over all time frames. In this case, we obtain $R^2 = 0.96$, which indicates a very strong correlation between the PD and GT primary spacing evolution.

\begin{table}[htbp]
\centering
\begin{tabular}{c c c c c} 
 \hline
 Simulation & Initial $T_{\textrm{interf}}$ & $G$ (K/$\mu$m) & $R^2$ & Acceleration\\
 \hline
 1 & -5.02 & 52.80 & 0.96 & \multirow{5}{*}{323.75} \\ 
 2 & -5.82 & 62.65 & 0.87 & \\ 
 3 & -5.22 & 35.08 & 0.98 & \\
 4 & -5.33 & 53.56 & 0.84 & \\
 5 & -5.67 & 48.16 & 0.94  & \\ 
 \hline
\end{tabular}
\caption{Prediction of the primary spacing during far-from-equilibrium solidification under different thermal conditions using the ADS. Simulation time: GT 28338.00 s vs. PD 87.53 s, which is measured excluding the time it takes for data outputs. 
}
\label{Table1}
\end{table}

We also evaluate the performance of the ADS under different thermal conditions to test the robustness of parametric interpolation. As summarized in Table~\ref{Table1}, the ADS accurately predicts dendritic growth and the evolution of primary spacing across a range of thermal gradients and interface temperatures, achieving $R^2 > 0.8$ for all simulations, which lies within the high-correlation regime. These results demonstrate that, once trained on a diverse set of thermal conditions, the ADS reliably generalizes to conditions not explicitly encountered during training.
Finally, we assess computational efficiency by comparing the simulation wall times between a high-performance GPU-accelerated PF solver implemented in CUDA and the ADS implemented in PyTorch, both running on the same Nvidia V100 GPUs. As shown in Table~\ref{Table1}, the ADS achieves a $323.75\times$ speed-up relative to direct PF simulations under the same conditions. The reported times correspond to the whole process each program is executed, excluding the data output and file writing stage. 
This dramatic acceleration highlights the potential of the ADS as a powerful surrogate for computationally intensive PF modeling. If we also consider the time required to generate the training dataset (comprising 24 training simulations, with approximately 1320 s needed to complete 4 simulations on a computing node with 4 V100 GPUs), the overall acceleration factor for completing all five extrapolative simulations in Table~\ref{Table1} using the ADS, compared to running all simulations with the PF solver, is approximately $16.95$, which already represents a significant speed-up.
While this benchmark is based on a limited set of simulations, it is important to note that as the number and size of extrapolative simulations increase, the effective acceleration becomes even more substantial. Since the ADS is scalable, and both data generation and training are one-time efforts, the associated overhead becomes negligible compared to the cumulative savings during inference, particularly in high-throughput studies.
These benchmark results demonstrate the potential of the ADS framework as an efficient and practical surrogate for microstructural predictions.

\section{Discussion}\label{sec13}

In this work, we introduced an autoregressive deep surrogate (ADS) framework based on a Scale-Invariant ConvNeXt architecture (SI-ConvNeXt) to accurately and efficiently predict dendritic microstructure evolution during alloy solidification. The framework is trained on short trajectories from quantitative PF simulations conducted in small domains and demonstrates excellent predictive capabilities when extrapolated to larger spatial and temporal scales, achieving a speed-up exceeding two orders of magnitude. This acceleration is achieved through three key mechanisms: (1) spatial downsampling, which reduces the resolution of training data while preserving essential interfacial dynamics; (2) larger time steps in the ADS predictions compared to the fine time stepping required in PF solvers; and (3) extrapolation in both spatial and temporal scales, enabling predictions over much larger domains and longer horizons than those used in training.
The ADS framework accurately captures key physical metrics of dendrite growth, including tip-selection constants, morphological symmetries, and the evolution of primary dendrite spacing. Our results show that the resolution-preserving design of the SI-ConvNeXt architecture is critical for modeling translationally invariant dynamics and enables robust extrapolation to large domain sizes in isothermal solidification, with extrapolation accuracy remaining consistent across the tested domain sizes.
Hyperparameter studies demonstrate that only a small training dataset is sufficient for the ADS to learn the microstructural dynamics, and that the ADS time step can be significantly larger than the PF time step.
Furthermore, by incorporating temperature fields, the framework generalizes to non-translationally invariant systems, such as directional solidification under imposed temperature gradients. These capabilities make the SI-ConvNeXt-based ADS framework particularly suitable for microstructural simulations at experimentally relevant scales. They also clarify practical considerations for scalable surrogate deployment and offer a promising pathway for integrating microstructure modeling into ICME workflows, thereby accelerating alloy design and optimization.

Additionally, this work introduces a complementary, data-driven approach for spatial coarsening in alloy solidification modeling. Unlike conventional methods such as the anti-trapping current \cite{Karma2001} or enhanced interfacial diffusivity \cite{ji_microstructural_2023}—which modify the physical model to accommodate upscaled interfaces and require rigorous mathematical analysis and additional corrective terms to mitigate thickness-dependent spurious interfacial effects—the present framework leverages high-fidelity PF simulations with physical interface thicknesses in small domains to generate a limited but accurate set of training data (the ADS can also be trained on PF simulation data with upscaled interfaces, if desired).
Although this high-fidelity data is subsequently coarsened via downsampling for surrogate model training, it preserves the physical fidelity of interfacial dynamics without altering interface properties. As a result, thickness-dependent spurious effects are avoided, and no additional corrective terms are needed. Moreover, since the coarsening is achieved through data reduction, the framework retains the potential to reconstruct high-resolution fields using data-driven and machine-learning approaches~\cite{du2024conditional,pan2023neural}.
Taken together, the presented ADS workflow provides a new, flexible pathway for accelerating microstructure simulations that is both complementary to and compatible with established PF modeling frameworks.

This work may be expanded in several directions. First, incorporating crystallographic orientation fields into the input and output channels of ADS would enable the extension of the ADS framework to polycrystalline systems, thereby allowing for the study of grain competition and grain boundary dynamics \cite{tourret2015growth,Mota2021,song_cell_2023,dorari2022growth} during alloy solidification.
Second, the SI-ConvNeXt architecture could be extended to three dimensions. This would enable efficient 3D microstructure predictions, particularly valuable for PF simulations where computational costs are often prohibitive.
Third, coupling the dynamical temperature field with the ADS framework would allow the model to capture latent-heat diffusion and other thermal effects beyond the frozen-temperature approximation. This capability could facilitate the quantitative modeling of interfacial dynamics in systems with strong thermal feedback~\cite{song_thermal-field_2018,ji2025phase}.
Fourth, thermodynamic and kinetic parameters, or complete CALPHAD-type free energy functions, could be incorporated into the ML models to inform microstructure predictions. This would broaden the predictive capabilities of the framework, enabling generalization beyond a single alloy system and facilitating its integration into ICME workflows for alloy design and optimization.
Lastly, the flexibility of the ADS framework allows it to be broadly applied to other pattern-forming phenomena in materials and energy systems where high-fidelity models face time and cost constraints. This positions ADS as a promising approach for accelerating microstructure modeling.

\section{Methods}\label{sec11}

\subsection*{Phase-field simulation of near-equilibrium solidification}
For modeling alloy solidification under near-equilibrium conditions, we employ a well-established quantitative PF model for dilute binary alloys~\cite{Karma2001,echebarria_quantitative_2004}. The model utilizes thin-interface asymptotics~\cite{karma_quantitative_1998,echebarria_quantitative_2004} to allow the diffuse interface thickness $W$ to be much larger than the capillary length $d_0$ while remaining quantitative. An anti-trapping current is incorporated to compensate for spurious solute trapping that would otherwise arise due to the increased $W$.
In this model, the evolution of the preconditioned phase field $\psi$ and the dimensionless supersaturation $U$ is governed by the following partial differential equations:
\begin{equation}
\begin{split} \label{PF_a}
a_s(\mathbf{n})^2 \frac{\partial \psi}{\partial t} =& a_s(\mathbf{n})^2 \left( \nabla^2 \psi - \phi \sqrt{2} |\vec{\nabla} \psi|^2 \right) + \vec{\nabla} \left[ a_s(\mathbf{n})^2 \right] \cdot \vec{\nabla} \psi \\
&+ \sum_{q}\left[\partial_{q}\left(|\vec{\nabla} \psi|^{2} a_{s}(\mathbf{n}) \frac{\partial a_{s}(\mathbf{n})}{\partial\left(\partial_{q} \psi\right)}\right)\right] + \sqrt{2} \left[ \phi -\lambda(1-\phi^2) U \right],
\end{split}
\end{equation}
and
\begin{equation}
\begin{split} \label{PF_b}
(1+k-(1-k) \phi) \frac{\partial U}{\partial t}=&\widetilde{D} \vec{\nabla} \cdot[(1-\phi) \vec{\nabla} U] + \vec{\nabla} \cdot\left[(1+(1-k) U) \frac{(1-\phi^2)}{2} \frac{\partial \psi}{\partial t} \frac{\vec{\nabla} \psi}{|\vec{\nabla} \psi|}\right] \\
&+[1+(1-k) U] \frac{\left(1-\phi^{2}\right)}{\sqrt{2}} \frac{\partial \psi}{\partial t},
\end{split}
\end{equation}
with
\begin{equation}
U=\frac{1}{1-k}\left[\frac{c / c_{l}^{0}}{(1-\phi) / 2+k(1+\phi) / 2}-1\right], \label{Supersaturation}
\end{equation}
where $c$ is the solute concentration field (in molar fraction), $c_l^0 = (T_m - T_0')/|m|$ is the equilibrium liquidus concentration at the reference temperature $T = T_0'$, $T_m$ is the melting temperature of the pure solvent, $|m|$ is the magnitude of liquidus slope, and $k=0.1$ is the solute partition coefficient. The summation over $q$ denotes spatial coordinates, i.e., $q = x, y$ in 2D. The dimensionless diffusivity is defined as
\[
\widetilde{D}=\frac{D \tau_{0}}{W^{2}}=a_{1} a_{2} \frac{W}{d_{0}},
\]
where $D$ is the solute diffusivity in the liquid, $\tau_0 = a_1 a_2 W^3 / (d_0 D)$ is the characteristic time scale, $a_1 = 5\sqrt{2}/8$, and $a_2 = 47/75$~\cite{karma_quantitative_1998}. The capillary length at the reference temperature $T_0$ is defined as
\[
d_0=\frac{\Gamma}{\Delta T_0}, \label{d_0}
\]
where $\Gamma$ is the Gibbs–Thomson coefficient and $\Delta T_0 = |m|(1-k)c_l^0$ is the freezing range. The coupling factor is defined as $\lambda = a_1 W / d_0$.
Under isothermal growth conditions, the reference temperature $T_0'$ is the fixed, which is below the liquidus temperature corresponding to the far-field solute concentration $c_\infty$, leading to an undercooling expressed through the supersaturation $\Omega = (c_l^0 - c_{\infty}) / [c_l^0 (1-k)]$. This modifies the definition of the capillary length to
\[
d_0^*=\frac{1-(1-k)\Omega}{k} d_0,
\]
which reduces to $d_0$ when $\Omega = 0$.
In the simulations, we consider cubic symmetry for the interfacial anisotropy with preferred growth directions aligned with the $x$ and $y$ axes: $a_s(\theta) = 1 + \epsilon_4 \cos{4\theta}$, where $\theta$ is the angle between the local surface normal and a fixed crystalline axis. The anisotropy strength is set to $\epsilon_4 = 0.007$.
No external noise is introduced during the simulations. The evolution equations~\eqref{PF_a}-\eqref{PF_b} are solved using a finite-difference method with explicit Euler time stepping on massively parallel graphic processing units (GPUs) using the computer unified device architecture (CUDA). The spatial discretization is $\Delta x / W = 0.8$ with $W / d_0^* = 5$, and the time step is $\Delta t / \tau_0 = 0.035$. Following Ref.~\cite{ji_isotropic_2022}, isotropic discretizations for leading differential operators such as the Laplacian and divergence are used. Periodic boundary conditions are applied on all sides.

For the small-domain datasets used to train and validate the surrogate, we conducted 25 PF simulations on $1024 \times 1024\, (d_0^*)^2$ domains (stored at $256 \times 256$ grid points and downsampled to $64 \times 64$ for learning). Each trajectory was initialized with 1-4 square-shaped nuclei, where the side length varied from $6$ to $24\, d_0^*$ and the centers were drawn uniformly at random (ensuring no overlap). A 20/5 split by trajectory was used for training and validation, respectively. Extrapolative simulations were performed on larger square domains with side length ranging from $1600$ to $8000\, d_0^*$, including: 1) multi-dendrite cases with eight square nuclei, each having independently sampled sizes and positions as described above; and 2) a single-dendrite case initialized with a circular nucleus of radius $40\, d_0^*$ (for side length $6400\, d_0^*$ only).

\subsection*{Phase-field simulation of far-from-equilibrium solidification}
For modeling alloy solidification under far-from-equilibrium conditions, we use a recent PF model for dilute binary alloys~\cite{ji_microstructural_2023,ji2025phase}. Details of this model, referred to as Model I, can be found in Ref.~\cite{ji2025phase}. This model intrinsically reproduces nonequilibrium interfacial effects at the solid-liquid interface using a physical interface thickness $W_0$ and employs enhanced solute diffusivity within the spatially diffuse interface region to accurately capture solute trapping with a larger interface width $W \equiv S W_0$, where $S$ is the upscaling ratio. Simulations with this model have successfully predicted dendritic-banding transitions and banded microstructures, in quantitative agreement with thin-film Al-Cu experiments~\cite{ji_microstructural_2023,ji2024microstructure}.
The evolution equations for the PF variable $\phi$ and the concentration field $c$ are derived variationally from the free-energy functional introduced in Ref.~\cite{Karma2003Phase-fieldFormation}, and are given by:
\begin{align}
\frac{a^2_s(\mathbf{n})}{a_k(\mathbf{n})} \frac{\partial \phi}{\partial t} =& \vec{\nabla} \cdot\left[a_s(\mathbf{n})^{2} \vec{\nabla} \phi\right] + \phi - \phi^{3} + \sum_{q}\left[\partial_{q}\left(|\vec{\nabla} \phi|^{2} a_s(\mathbf{n}) \frac{\partial a_s(\mathbf{n})}{\partial\left(\partial_{q} \phi\right)}\right)\right] \label{Dimensionless_p} \\
&- \tilde{\lambda} g^{\prime}(\phi) \left[\tilde{c} + \widetilde{T} e^{b(1+g(\phi))} \right], \nonumber \\
\frac{\partial \tilde{c}}{\partial t} =& \widetilde{D}_l \vec{\nabla} \cdot \left \{ q(\phi) \tilde{c} \vec{\nabla}[\ln \tilde{c}-b g(\phi)] \right \}, \label{Dimensionless_c}
\end{align}
where $\tilde{\lambda} \equiv \lambda c_\infty$, $\tilde{c} \equiv c / c_\infty$, and $\widetilde{D}_l \equiv D_l \tau_0 / W^2 = D_l / (\Gamma \mu_k^0)$ are dimensionless quantities, with $c_\infty$ the nominal alloy concentration, $D_l$ the liquid solute diffusivity, $\Gamma = \gamma_0 T_M / L$ the Gibbs-Thomson coefficient ($\gamma_0$ is the interfacial free energy, $T_M$ the melting temperature of the pure solvent, and $L$ the latent heat of fusion per volume), and $\mu_k^0$ the kinetic coefficient. The parameters are related via $b \equiv \ln k_e / 2 < 0$ (with $k_e$ the equilibrium partition coefficient), $\tau_0 = (S W_0)^2 / (\Gamma \mu_k^0)$ the characteristic time scale, and $\lambda = a_1^0 b m_e S W_0 / [\Gamma (k_e - 1)] > 0$, where $m_e > 0$ is the equilibrium liquidus slope and $a_1^0 = 2\sqrt{2}/3$.
The interpolation functions are defined as $g(\phi) = 15(\phi - 2\phi^3/3 + \phi^5/5)/8$ and $q(\phi) = (1 - \phi)/2$ for $S=1$ (for $S > 1$, $q(\phi)$ is a nonlinear function of $\phi$). The anisotropic interface width and time constant are defined as $W(\mathbf{n}) = S W_0 a_s(\mathbf{n})$ and $\tau(\mathbf{n}) = \tau_0 a_s^2(\mathbf{n}) / a_k(\mathbf{n})$, respectively. These expressions model the anisotropic interfacial free energy $\gamma(\mathbf{n}) = \gamma_0 a_s(\mathbf{n})$ and kinetic coefficient $\mu_k(\mathbf{n}) = \mu_k^0 a_k(\mathbf{n})$, where $\mathbf{n}$ denotes the interface normal. The index $q$ represents spatial coordinates, $q = x,\, y$.
For directional solidification under the frozen temperature approximation (FTA), the dimensionless temperature field is given by:
\begin{equation}
\widetilde{T} = \frac{T - T_M}{m_e c_\infty} = \frac{x-x_0-\widetilde{V}_{\mathrm{iso}} t}{\widetilde{l}_T}-1,
\end{equation}
where $\widetilde{V}_{\mathrm{iso}} \equiv \tau_0 V_{\mathrm{iso}} / W$ and $\widetilde{l}_T \equiv l_T / W = (m_e c_\infty / G) / W$ denote the dimensionless pulling speed and thermal length, respectively, $G$ is the temperature gradient, and $V_{\mathrm{iso}}$ is the isotherm velocity. The reference position $x_0$ corresponds to the location where $T = T_L - m_e c_\infty$, with $T_L$ the liquidus temperature at $c_\infty$.
In the simulations, we consider a cubic symmetry for both the interface free energy anisotropy $a_s(\theta) = 1+\epsilon_s \cos{4\theta}$, and the kinetic anisotropy $a_k(\theta) = 1+\epsilon_k \cos{4\theta}$, with the anisotropy strength $\epsilon_s=0.012$ and $\epsilon_k=0.1$. In our simulations, the preferred growth directions are aligned with the $x$ and $y$ axes. Similar to PF simulations of near-equilibrium solidification, we solve Eqs.~\eqref{Dimensionless_p}-\eqref{Dimensionless_c} using the finite-difference and Euler explicit time stepping on GPUs, with leading differential terms discretized by isotropic schemes~\cite{ji_isotropic_2022}, and $\Delta x/W=0.8$ and $\Delta t/\tau_0=0.096$. All simulations consider a physical interface thickness, i.e., $W=W_0$. Zero-flux (Neumann) boundary conditions are applied in the transverse ($y$) direction, while the domain is simulated in a moving frame along the growth ($x$) direction. The materials parameters for simulations of the Al-Cu alloys are listed in Table I of Ref.~\cite{ji2025phase}.

The small-domain training dataset comprised $N_{\mathrm{train}} = 24$ PF simulations conducted in a domain of size $L_x \times L_y = 1.28 \times 0.32\, \mu\mathrm{m}^2$. Each simulation was initialized from a one-dimensional steady-state planar interface at $V_{\mathrm{iso}} = 0.12~\mathrm{m/s}$ and perturbed by a single sinusoidal wave with wavelength $\lambda \in [0.08,\, 0.32]~\mu\mathrm{m}$ (chosen uniformly at random). Thermal gradients and interface temperatures were independently and uniformly sampled as $G \in [10,\, 100]~\mathrm{K/\mu m}$ and $T_{\mathrm{interf}} \in [-6,\, -5]$ (in normalized units), respectively. An additional four simulations were performed for validation.
Extrapolative simulations were performed in a larger domain of $L_x \times L_y = 5.12 \times 0.64\, \mu\mathrm{m}^2$ with composite perturbations generated as linear combinations of multiple sinusoidal components. The test thermal conditions $(G,\, T_{\mathrm{interf}})$ were sampled from the same ranges but were held out during training.

\subsection*{Machine-learning model}

We investigate the complex spatiotemporal evolution of state variables $\phi$ and $u$ by autoregressive deep surrogate (ADS) models $\mathcal{F}$:
\begin{equation}
(\phi_{n+1}, u_{n+1}) = \mathcal{F}(\phi_n, u_n; T_n),
\end{equation}
where $u$ denotes either the supersaturation field $U$ (for isothermal dendrite growth) or the concentration field $c$ (for directional solidification), and $T_n$ is the optional time-dependent temperature field for directional solidification. We construct our surrogate model leveraging the backbone of ConvNeXt~\cite{liu2022convnet}, modified to refrain from using a bottleneck architecture, thereby maintaining resolution throughout the forward pass to facilitate spatial extrapolation. This design yields a Scale-Invariant ConvNeXt (SI-ConvNeXt) architecture. As shown in Fig.~\ref{fig:diagram}(a), the scale-invariant convolutional block consists of a depth-wise convolution followed by a point-wise convolution, with the spatial dimensions remaining unchanged during these operations. A residual connection is employed to mitigate gradient vanishing and exploding issues~\cite{he2016deep}. The forward process inside a convolutional block is:
\begin{equation}
\mathbf{h}_{l+1} = \mathbf{h}_l + \sigma\bigg(\mathcal{P}\big(
\mathcal{C}(\mathbf{h}_l)\big)\bigg),
\end{equation}
where $\mathcal{C}$ is the depth-wise convolution with weight matrix $\mathbf{W}^{(c)}_{k\times k}$ defined by kernel size $k=3$ and channel index $c$, $\mathcal{P}$ defines point-wise scalar weights $\mathbf{W}^{p}_{1\times 1}$, and $\sigma$ is the non-linear activation function. We use standard layer normalization techniques, which are prevalent in transformer architectures; the mathematical expressions are omitted here for clarity. A learnable stem layer is used to extract features from the input state variables $(\phi_n, u_n; T_n)$, and a learnable head layer maps the predicted features to the next-step variables $(\phi_{n+1}, u_{n+1})$.

Next-step prediction is straightforward to implement but can accumulate errors over time. To mitigate this, we introduce noise injection during training, which adds a small amount of noise to the input to enhance robustness for imperfect inputs encountered during long rollouts. In the current work, artificial Gaussian noise is added to the inputs before feeding them into the network:
\begin{align}
\tilde{\phi}_n &= \phi_n^{\mathrm{gt}} + \epsilon_\phi, \quad \epsilon_\phi \sim \mathcal{N}(0, \sigma_\phi^2), \\
\tilde{u}_n &= u_n^{\mathrm{gt}} + \epsilon_u, \quad \epsilon_u \sim \mathcal{N}(0, \sigma_u^2),
\end{align}
where $\sigma_\phi=0.001$ and $\sigma_u=0.001$ are hyperparameters setting the standard deviations of the zero-mean Gaussian noise added to $\phi$ and $u$ during training. The loss function is defined as:
\begin{align}
\mathcal{L} &= \mathbb{E}\left[\|(\hat{\boldsymbol{\phi}}_{n+1},\hat{\mathbf{u}}_{n+1}) - ( \boldsymbol{\phi}_{n+1}^{\mathrm{gt}},\mathbf{u}_{n+1}^{\mathrm{gt}})\|^2\right],
\end{align}
where $(\hat{\phi}_{n+1},\hat{u}_{n+1}) = \mathcal{F}(\tilde{\phi}_n,\tilde{u}_n; T_n)$. During training, the temperature field (when applicable) is predefined and concatenated as an additional input channel but excluded from the loss calculation.

Unlike training, which uses paired state variables $(\phi_n,u_n)$, testing starts only with initial conditions $(\phi_0,u_0)$. We iteratively apply the learned surrogate model to predict trajectories as:
\begin{align}
\left(\hat{\phi}_N,\hat{u}_N\right) &= \mathcal{F}_{T_{N-1}} \circ \mathcal{F}_{T_{N-2}} \circ \cdots \circ \mathcal{F}_{T_0} (\phi_0,u_0),
\end{align}
where $\mathcal{F}_{T_k}(\cdots) \equiv \mathcal{F}(\cdots; T_k)$. For directional solidification, the predicted temperature output is discarded, and the predefined $T_{n+1}$ is concatenated for the next step.

In addition to noise injection, data augmentation is an effective technique for improving validation performance. In the present work, we apply the 2D $4m$ point group symmetry to the training datasets for isothermal dendrite growth.
We index the discrete fields on a uniform $N_x \times N_y$ grid by integers $i \in \{1, \dots, N_x\}$ and $j \in \{1, \dots, N_y\}$. We also define the reflected indices as $\bar{i} = N_x + 1 - i$ and $\bar{j} = N_y + 1 - j$. The symmetry operations transform the state variables as follows:
\begin{align*}
&(\phi_{ij}, u_{ij}), \quad
(\phi_{\bar{i}j}, u_{\bar{i}j}), \quad
(\phi_{i\bar{j}}, u_{i\bar{j}}), \quad
(\phi_{\bar{i}\bar{j}}, u_{\bar{i}\bar{j}}), \\
&(\phi_{ji}, u_{ji}), \quad
(\phi_{\bar{j}\bar{i}}, u_{\bar{j}\bar{i}}), \quad
(\phi_{j\bar{i}}, u_{j\bar{i}}), \quad
(\phi_{\bar{j}i}, u_{\bar{j}i}).
\end{align*}
Here, the first row corresponds to the identity, reflections about the $x$- and $y$-axes, and a $180^\circ$ rotation. The second row represents reflections about the two diagonals, as well as $90^\circ$ rotations. This augmentation is not applied to directional solidification datasets, as the fixed temperature gradient direction breaks rotational invariance.

The ML model is implemented using the Neural Phase Simulation (NPS) package \cite{yang2021self,bertin2023accelerating}. The model is trained for 1000 epochs using $M=7$ convolutional blocks. We employ the AdamW optimizer~\cite{loshchilov2017decoupled} with weight decay for improved generalization. To optimize the learning rate, we trained four surrogate models per dataset with learning rates ranging from $2 \times 10^{-4}$ to $8 \times 10^{-4}$ in increments of $2 \times 10^{-4}$, selecting the model with the lowest mean squared error on a validation simulation. During convolutional operations, periodic padding is used for isothermal cases to match periodic boundary conditions, while zero-flux (reflective) padding is used for directional solidification to match Neumann boundary conditions employed in the numerical simulations.

\subsection*{Metrics for the error of predicted fields}
For each field $i\in\{\phi, u\}$, where $u$ denotes the supersaturation field $U$ for isothermal dendrite growth or the concentration field $c$ for directional solidification, we measure the point-wise prediction error with the spatially averaged mean-squared error (MSE). At every stored time step, we compute
\begin{equation}
\mathrm{MSE}^i(t)=
\frac{1}{r_i^{2}N_xN_y}
\sum_{x=1}^{N_x}\sum_{y=1}^{N_y}
\bigl[\hat{f}^{\,i}(t,x,y)-f^{\,i}(t,x,y)\bigr]^2,
\label{eq:mse_inst}
\end{equation}
with $r_i=\max f^{\,i}-\min f^{\,i}$ taken from the initial ground-truth field. The single score reported is the temporal average
\begin{equation}
\bigl\langle\mathrm{MSE}^i\bigr\rangle_t =
T^{-1}\sum_{n}\mathrm{MSE}^i(t)\,\Delta t,
\end{equation}
evaluated with Simpson’s rule over the prediction horizon $T$.

\subsection*{Metrics for evaluating dendrite symmetry}
For the isothermal dendrite growth case, an ideal dendrite morphology is unchanged by quarter-turn rotations. We therefore measure the departure from four-fold symmetry in the predicted fields with the time-averaged error
\begin{align}
\langle E_{\mathrm{sym}}^{\,i}\rangle_t
&= \frac{1}{3T}
  \sum_{n=0}^{N_t-1}\!\Delta t
  \sum_{\theta\in\{90^\circ,180^\circ,270^\circ\}}
  \frac{1}{N_xN_y}
  \sum_{x=1}^{N_x}\sum_{y=1}^{N_y}
  {} \nonumber \\
&\quad\times
  \bigl[\hat{f}^{\,i}(t,x,y)-\mathcal R_{\theta}\hat{f}^{\,i}(t,x,y)\bigr]^{2}\,,
\end{align}
where $\hat{f}^{\,i}$ is the predicted field ($\phi$ or $U$), $\mathcal R_{\theta}$ rotates it by $\theta$ around the domain center, $\Delta t$ is the physical time step, and $T=\sum_{n}\Delta t$ is the prediction horizon. The metric equals zero for perfect four-fold rotational symmetry and grows as the pattern becomes less symmetric.

\backmatter

\bmhead{Acknowledgements}

This work was performed under the auspices of the U.S. Department of Energy by Lawrence Livermore National Laboratory (LLNL) under contract DE-AC52-07NA27344. The work of KJ and TWH was funded by the Laboratory Directed Research and Development (LDRD) Program at LLNL under a project with a tracking code 23-SI-002. The work of LS and FZ was funded by the LDRD Program at LLNL under a project with a tracking code 25-ERD-002. Computing support for this work came from LLNL Institutional Computing Grand Challenge program.

\section*{Declarations}

\subsection*{Data availability}

The datasets used and analyzed during the current study are available from the corresponding author upon reasonable request.

\subsection*{Code availability}

The NPS package is publicly available at \url{https://github.com/Critical-Materials-Institute/NPS}. Other codes and scripts used in this study are available from the corresponding author upon reasonable request.

\subsection*{Author contribution}

KJ: Conceptualization, Investigation, Methodology, Resources, Software, Validation, Visualization, Writing – original draft, Writing – review \& editing; LS: Investigation, Methodology, Software, Validation, Visualization, Writing – original draft, Writing – review \& editing; SL: Validation, Writing – review \& editing; FZ: Funding acquisition, Investigation, Methodology, Software, Validation, Writing – review \& editing; TWH: Conceptualization, Funding acquisition, Validation, Writing – review \& editing

\subsection*{Competing interests}

The authors declare no competing interests.

\bibliography{sn-bibliography}

@article{bertin2024learning,
  title={Learning dislocation dynamics mobility laws from large-scale MD simulations},
  author={Bertin, Nicolas and Bulatov, Vasily V and Zhou, Fei},
  journal={npj Computational Materials},
  volume={10},
  number={1},
  pages={192},
  year={2024},
  publisher={Nature Publishing Group UK London}
}

@article{fan2024accelerate,
  title={Accelerate microstructure evolution simulation using graph neural networks with adaptive spatiotemporal resolution},
  author={Fan, Shaoxun and Hitt, Andrew L and Tang, Ming and Sadigh, Babak and Zhou, Fei},
  journal={Machine Learning: Science and Technology},
  volume={5},
  number={2},
  pages={025027},
  year={2024},
  publisher={IOP Publishing}
}

@article{tourret2023morphological,
  title={Morphological stability of solid-liquid interfaces under additive manufacturing conditions},
  author={Tourret, Damien and Klemm-Toole, Jonah and Castellanos, Adriana Eres and Rodgers, Brian and Becker, Gus and Saville, Alec and Ellyson, Ben and Johnson, Chloe and Milligan, Brian and Copley, John and others},
  journal={Acta Materialia},
  volume={250},
  pages={118858},
  year={2023},
  publisher={Elsevier}
}

@article{trivedi1980theory,
  title={Theory of dendritic growth during the directional solidification of binary alloys},
  author={Trivedi, R},
  journal={Journal of crystal Growth},
  volume={49},
  number={2},
  pages={219--232},
  year={1980},
  publisher={Elsevier}
}

@article{rickman2019materials,
  title={Materials informatics: From the atomic-level to the continuum},
  author={Rickman, Jeffrey M and Lookman, Turab and Kalinin, Sergei V},
  journal={Acta Materialia},
  volume={168},
  pages={473--510},
  year={2019},
  publisher={Elsevier}
}

@article{dorari2022growth,
  title={Growth competition between columnar dendritic grains--The role of microstructural length scales},
  author={Dorari, Elaheh and Ji, Kaihua and Guillemot, Gildas and Gandin, Charles-Andr{\'e} and Karma, Alain},
  journal={Acta Materialia},
  volume={223},
  pages={117395},
  year={2022},
  publisher={Elsevier}
}

@article{konig2023solidification,
  title={Solidification modes during additive manufacturing of steel revealed by high-speed X-ray diffraction},
  author={K{\"o}nig, Hans-Henrik and Pettersson, Niklas Holl{\"a}nder and Durga, A and Van Petegem, Steven and Grolimund, Daniel and Chuang, Andrew Chihpin and Guo, Qilin and Chen, Lianyi and Oikonomou, Christos and Zhang, Fan and others},
  journal={Acta Materialia},
  volume={246},
  pages={118713},
  year={2023},
  publisher={Elsevier}
}

@article{losert1998evolution,
  title={Evolution of dendritic patterns during alloy solidification: Onset of the initial instability},
  author={Losert, W and Shi, BQ and Cummins, HZ},
  journal={Proceedings of the National Academy of Sciences},
  volume={95},
  number={2},
  pages={431--438},
  year={1998},
  publisher={The National Academy of Sciences}
}

@article{utter2005double,
  title={Double dendrite growth in solidification},
  author={Utter, Brian and Bodenschatz, Eberhard},
  journal={Physical Review E—Statistical, Nonlinear, and Soft Matter Physics},
  volume={72},
  number={1},
  pages={011601},
  year={2005},
  publisher={APS}
}

@incollection{Karma2003Phase-fieldFormation,
    title = {{Phase-field models of microstructural pattern formation}},
    year = {2003},
    booktitle = {Thermodynamics, Microstructures and Plasticity},
    author = {Karma, A.},
    editor = {Finel, A. and Mazi{\`{e}}re, D. and Veron, M.},
    pages = {65--89},
    publisher = {Springer},
    url = {https://link.springer.com/book/9781402013676},
    address = {Dordrecht},
    isbn = {978-1-4020-1367-6}
}

@article{boettinger2000solidification,
  title={Solidification microstructures: recent developments, future directions},
  author={Boettinger, William J and Coriell, Sam R and Greer, AL and Karma, A and Kurz, W and Rappaz, M and Trivedi, R},
  journal={Acta materialia},
  volume={48},
  number={1},
  pages={43--70},
  year={2000},
  publisher={Elsevier}
}

@article{glasner2001nonlinear,
  title={Nonlinear preconditioning for diffuse interfaces},
  author={Glasner, Karl},
  journal={Journal of Computational Physics},
  volume={174},
  number={2},
  pages={695--711},
  year={2001},
  publisher={Elsevier}
}

@article{li2012dendrite,
  title={Dendrite fragmentation and columnar-to-equiaxed transition during directional solidification at lower growth speed under a strong magnetic field},
  author={Li, Xi and Gagnoud, Annie and Fautrelle, Yves and Ren, Zhongming and Moreau, Rene and Zhang, Yudong and Esling, Claude},
  journal={Acta materialia},
  volume={60},
  number={8},
  pages={3321--3332},
  year={2012},
  publisher={Elsevier}
}

@article{bonneville2025accelerating,
  title={Accelerating phase field simulations through a hybrid adaptive Fourier neural operator with U-net backbone},
  author={Bonneville, Christophe and Bieberdorf, Nathan and Hegde, Arun and Asta, Mark and Najm, Habib N and Capolungo, Laurent and Safta, Cosmin},
  journal={npj Computational Materials},
  volume={11},
  number={1},
  pages={14},
  year={2025},
  publisher={Nature Publishing Group UK London}
}

@article{tourret2015oscillatory,
  title={Oscillatory cellular patterns in three-dimensional directional solidification},
  author={Tourret, Damien and Debierre, J-M and Song, Younggil and Mota, Fatima L and Bergeon, Nathalie and Guerin, Rahma and Trivedi, Rohit and Billia, Bernard and Karma, Alain},
  journal={Physical Review E},
  volume={92},
  number={4},
  pages={042401},
  year={2015},
  publisher={APS}
}

@article{tseng2023deep,
  title={Deep learning model to predict ice crystal growth},
  author={Tseng, Bor-Yann and Guo, Chen-Wei Conan and Chien, Yu-Chen and Wang, Jyn-Ping and Yu, Chi-Hua},
  journal={Advanced Science},
  volume={10},
  number={21},
  pages={2207731},
  year={2023},
  publisher={Wiley Online Library}
}

@article{bertin2023accelerating,
  title={Accelerating discrete dislocation dynamics simulations with graph neural networks},
  author={Bertin, Nicolas and Zhou, Fei},
  journal={Journal of Computational Physics},
  volume={487},
  pages={112180},
  year={2023},
  publisher={Elsevier}
}

@article{montes2021accelerating,
  title={Accelerating phase-field-based microstructure evolution predictions via surrogate models trained by machine learning methods},
  author={Montes de Oca Zapiain, David and Stewart, James A and Dingreville, R{\'e}mi},
  journal={npj Computational Materials},
  volume={7},
  number={1},
  pages={3},
  year={2021},
  publisher={Nature Publishing Group UK London}
}

@article{yang2021self,
  title={Self-supervised learning and prediction of microstructure evolution with convolutional recurrent neural networks},
  author={Yang, Kaiqi and Cao, Yifan and Zhang, Youtian and Fan, Shaoxun and Tang, Ming and Aberg, Daniel and Sadigh, Babak and Zhou, Fei},
  journal={Patterns},
  volume={2},
  number={5},
  year={2021},
  publisher={Elsevier}
}

@article{zhang2023quantitative,
  title={Quantitative phase field model for electrochemical systems},
  author={Zhang, Jin and Chadwick, Alexander F and Voorhees, Peter W},
  journal={Journal of The Electrochemical Society},
  volume={170},
  number={12},
  pages={120503},
  year={2023},
  publisher={IOP Publishing}
}

@article{chen2015modulation,
  title={Modulation of dendritic patterns during electrodeposition: A nonlinear phase-field model},
  author={Chen, Lei and Zhang, Hao Wei and Liang, Lin Yun and Liu, Zhe and Qi, Yue and Lu, Peng and Chen, James and Chen, Long-Qing},
  journal={Journal of Power Sources},
  volume={300},
  pages={376--385},
  year={2015},
  publisher={Elsevier}
}

@article{bai2016transition,
  title={Transition of lithium growth mechanisms in liquid electrolytes},
  author={Bai, Peng and Li, Ju and Brushett, Fikile R and Bazant, Martin Z},
  journal={Energy \& Environmental Science},
  volume={9},
  number={10},
  pages={3221--3229},
  year={2016},
  publisher={Royal Society of Chemistry}
}

@article{chen2002phase,
  title={Phase-field models for microstructure evolution},
  author={Chen, Long-Qing},
  journal={Annual review of materials research},
  volume={32},
  number={1},
  pages={113--140},
  year={2002},
  publisher={Annual Reviews 4139 El Camino Way, PO Box 10139, Palo Alto, CA 94303-0139, USA}
}

@article{ramana2009study,
  title={Study of the Li-insertion/extraction process in LiFePO4/FePO4},
  author={Ramana, CV and Mauger, A and Gendron, F and Julien, CM and Zaghib, K},
  journal={Journal of Power Sources},
  volume={187},
  number={2},
  pages={555--564},
  year={2009},
  publisher={Elsevier}
}

@article{geslin2015topology,
  title={Topology-generating interfacial pattern formation during liquid metal dealloying},
  author={Geslin, Pierre-Antoine and McCue, Ian and Gaskey, Bernard and Erlebacher, Jonah and Karma, Alain},
  journal={Nature communications},
  volume={6},
  number={1},
  pages={8887},
  year={2015},
  publisher={Nature Publishing Group UK London}
}

@article{zhao2023understanding,
  title={Understanding and design of metallic alloys guided by phase-field simulations},
  author={Zhao, Yuhong},
  journal={npj Computational Materials},
  volume={9},
  number={1},
  pages={94},
  year={2023},
  publisher={Nature Publishing Group UK London}
}

@article{ji2025phase,
  title = {Phase-field model of alloy solidification far from chemical equilibrium at the solid-liquid interface},
  author = {Ji, Kaihua and Zhong, Mingwang and Karma, Alain},
  journal = {Phys. Rev. Res.},
  pages = {--},
  year = {2025},
  month = {Jul},
  publisher = {American Physical Society},
  doi = {10.1103/vyd6-nj4h},
  url = {https://link.aps.org/doi/10.1103/vyd6-nj4h}
}

@article{tourret2015growth,
  title={Growth competition of columnar dendritic grains: A phase-field study},
  author={Tourret, Damien and Karma, Alain},
  journal={Acta Materialia},
  volume={82},
  pages={64--83},
  year={2015},
  doi={10.1016/j.actamat.2014.08.049},
  publisher={Elsevier}
}

@article{ji2024microstructure,
  title={Microstructure development during rapid alloy solidification},
  author={Ji, Kaihua and Clarke, Amy J. and McKeown, Joseph T. and Karma, Alain},
  journal={MRS Bulletin},
  doi={10.1557/s43577-024-00717-6},
  year={2024}
}

@article{karma_quantitative_1998,
	title = {Quantitative phase-field modeling of dendritic growth in two and three dimensions},
	volume = {57},
	issn = {1063-651X, 1095-3787},
	url = {https://link.aps.org/doi/10.1103/PhysRevE.57.4323},
	doi = {10.1103/PhysRevE.57.4323},
	number = {4},
	urldate = {2023-11-20},
	journal = {Physical Review E},
	author = {Karma, Alain and Rappel, Wouter-Jan},
	month = apr,
	year = {1998},
	pages = {4323--4349},
}

@article{echebarria_quantitative_2004,
	title = {Quantitative phase-field model of alloy solidification},
	volume = {70},
	issn = {1539-3755, 1550-2376},
	url = {https://link.aps.org/doi/10.1103/PhysRevE.70.061604},
	doi = {10.1103/PhysRevE.70.061604},
	number = {6},
	journal = {Physical Review E},
	author = {Echebarria, Blas and Folch, Roger and Karma, Alain and Plapp, Mathis},
	year = {2004},
	pages = {061604},
}

@article{karma_atomistic_2016,
	title = {Atomistic to continuum modeling of solidification microstructures},
	volume = {20},
	issn = {13590286},
	url = {https://linkinghub.elsevier.com/retrieve/pii/S1359028615300061},
	doi = {10.1016/j.cossms.2015.09.001},
	number = {1},
	urldate = {2023-11-20},
	journal = {Current Opinion in Solid State and Materials Science},
	author = {Karma, Alain and Tourret, Damien},
	month = feb,
	year = {2016},
	pages = {25--36},
}

@article{clarke_microstructure_2017,
	title = {Microstructure selection in thin-sample directional solidification of an {Al}-{Cu} alloy: {In} situ {X}-ray imaging and phase-field simulations},
	volume = {129},
	issn = {13596454},
	shorttitle = {Microstructure selection in thin-sample directional solidification of an {Al}-{Cu} alloy},
	url = {https://linkinghub.elsevier.com/retrieve/pii/S1359645417301489},
	doi = {10.1016/j.actamat.2017.02.047},
	urldate = {2023-11-20},
	journal = {Acta Materialia},
	author = {Clarke, A.J. and Tourret, D. and Song, Y. and Imhoff, S.D. and Gibbs, P.J. and Gibbs, J.W. and Fezzaa, K. and Karma, A.},
	month = may,
	year = {2017},
	pages = {203--216},
}

@article{ji_microstructural_2023,
	title = {Microstructural {Pattern} {Formation} during {Far}-from-{Equilibrium} {Alloy} {Solidification}},
	volume = {130},
	url = {https://link.aps.org/doi/10.1103/PhysRevLett.130.026203},
	doi = {10.1103/PhysRevLett.130.026203},
	number = {2},
	urldate = {2023-10-06},
	journal = {Physical Review Letters},
	author = {Ji, Kaihua and Dorari, Elaheh and Clarke, Amy J. and Karma, Alain},
	month = jan,
	year = {2023},
	pages = {026203},
}

@article{song_thermal-field_2018,
	title = {Thermal-field effects on interface dynamics and microstructure selection during alloy directional solidification},
	volume = {150},
	issn = {13596454},
	doi = {10.1016/j.actamat.2018.03.012},
	urldate = {2021-03-08},
	journal = {Acta Materialia},
	author = {Song, Y. and Tourret, D. and Mota, F. L. and Pereda, J. and Billia, B. and Bergeon, N. and Trivedi, R. and Karma, A.},
	month = may,
	year = {2018},
	pages = {139--152},
}

@article{ji_isotropic_2022,
	title = {Isotropic finite-difference approximations for phase-field simulations of polycrystalline alloy solidification},
	volume = {457},
	issn = {0021-9991},
	doi = {10.1016/J.JCP.2022.111069},
	urldate = {2022-07-17},
	journal = {Journal of Computational Physics},
	author = {Ji, Kaihua and Tabrizi, Amirhossein Molavi and Karma, Alain},
	month = may,
	year = {2022},
	pages = {111069},
}

@article{Karma2001,
	title = {Phase-field formulation for quantitative modeling of alloy solidification},
	volume = {87},
	issn = {10797114},
	doi = {10.1103/PhysRevLett.87.115701},
	number = {11},
	journal = {Physical Review Letters},
	author = {Karma, Alain},
	year = {2001},
	pages = {115701},
}

@article{tourret_phase-field_2022,
	title = {Phase-field modeling of microstructure evolution: {Recent} applications, perspectives and challenges},
	volume = {123},
	issn = {0079-6425},
	doi = {10.1016/J.PMATSCI.2021.100810},
	urldate = {2022-03-19},
	journal = {Progress in Materials Science},
	author = {Tourret, Damien and Liu, Hong and LLorca, Javier},
	month = jan,
	year = {2022},
	pages = {100810},
}

@article{song_cell_2023,
	title = {Cell invasion during competitive growth of polycrystalline solidification patterns},
	volume = {14},
	copyright = {2023 The Author(s)},
	issn = {2041-1723},
	url = {https://www.nature.com/articles/s41467-023-37458-0},
	doi = {10.1038/s41467-023-37458-0},
	number = {1},
	urldate = {2023-10-12},
	journal = {Nature Communications},
	author = {Song, Younggil and Mota, Fatima L. and Tourret, Damien and Ji, Kaihua and Billia, Bernard and Trivedi, Rohit and Bergeon, Nathalie and Karma, Alain},
	month = apr,
	year = {2023},
	pages = {2244},
}

@book{kurz1989fundamentals,
	title = {Fundamentals of solidification},
	publisher = {Trans Tech Publications},
	author = {Kurz, Wilfried and Fisher, David J},
        address = {Aedermannsdorf},
	year = {1989},
}

@article{boettinger_phase-field_2002,
	title = {Phase-field simulation of solidification},
	volume = {32},
	issn = {00846600},
	doi = {10.1146/annurev.matsci.32.101901.155803},
	journal = {Annual Review of Materials Science},
	author = {Boettinger, W. J. and Warren, James A. and Beckermann, C. and Karma, A.},
	year = {2002},
	pages = {163--194},
}

@book{dantzig_solidification_2016,
	title = {Solidification},
	isbn = {2-940222-97-5},
	publisher = {EPFL press},
	author = {Dantzig, Jonathan A and Rappaz, Michel},
        address = {Lausanne},
	year = {2016},
}

@article{barbieri_predictions_1989,
    title = {Predictions of dendritic growth rates in the linearized solvability theory},
    volume = {39},
    issn = {10502947},
    url = {https://journals.aps.org/pra/abstract/10.1103/PhysRevA.39.5314},
    doi = {10.1103/PhysRevA.39.5314},
    number = {10},
    urldate = {2021-05-24},
    journal = {Physical Review A},
    author = {Barbieri, A. and Langer, J. S.},
    month = may,
    year = {1989},
    pages = {5314--5325},
}

@article{Karma2000,
    title = {Three-dimensional dendrite-tip morphology at low undercooling},
    volume = {61},
    issn = {1063651X},
    doi = {10.1103/PhysRevE.61.3996},
    number = {4},
    journal = {Physical Review E},
    author = {Karma, Alain and Lee, Youngyih H. and Plapp, Mathis},
    year = {2000},
}

@article{li2020fourier,
  title={Fourier neural operator for parametric partial differential equations},
  author={Li, Zongyi and Kovachki, Nikola and Azizzadenesheli, Kamyar and Liu, Burigede and Bhattacharya, Kaushik and Stuart, Andrew and Anandkumar, Anima},
  journal={arXiv preprint arXiv:2010.08895},
  year={2020}
}

@article{lu2021learning,
  title={Learning nonlinear operators via DeepONet based on the universal approximation theorem of operators},
  author={Lu, Lu and Jin, Pengzhan and Pang, Guofei and Zhang, Zhongqiang and Karniadakis, George Em},
  journal={Nature machine intelligence},
  volume={3},
  number={3},
  pages={218--229},
  year={2021},
  publisher={Nature Publishing Group UK London}
}

@article{pan2023neural,
  title={Neural implicit flow: a mesh-agnostic dimensionality reduction paradigm of spatio-temporal data},
  author={Pan, Shaowu and Brunton, Steven L and Kutz, J Nathan},
  journal={Journal of Machine Learning Research},
  volume={24},
  number={41},
  pages={1--60},
  year={2023}
}

@article{du2024conditional,
  title={Conditional neural field latent diffusion model for generating spatiotemporal turbulence},
  author={Du, Pan and Parikh, Meet Hemant and Fan, Xiantao and Liu, Xin-Yang and Wang, Jian-Xun},
  journal={Nature Communications},
  volume={15},
  number={1},
  pages={10416},
  year={2024},
  publisher={Nature Publishing Group UK London}
}

@inproceedings{pfaff2020learning,
  title={Learning mesh-based simulation with graph networks},
  author={Pfaff, Tobias and Fortunato, Meire and Sanchez-Gonzalez, Alvaro and Battaglia, Peter},
  booktitle={International conference on learning representations},
  year={2020}
}

@article{han2022predicting,
  title={Predicting physics in mesh-reduced space with temporal attention},
  author={Han, Xu and Gao, Han and Pfaff, Tobias and Wang, Jian-Xun and Liu, Li-Ping},
  journal={arXiv preprint arXiv:2201.09113},
  year={2022}
}

@article{sun2023unifying,
  title={Unifying predictions of deterministic and stochastic physics in mesh-reduced space with sequential flow generative model},
  author={Sun, Luning and Han, Xu and Gao, Han and Wang, Jian-Xun and Liu, Liping},
  journal={Advances in Neural Information Processing Systems},
  volume={36},
  pages={60636--60660},
  year={2023}
}

@article{gao2024bayesian,
  title={Bayesian conditional diffusion models for versatile spatiotemporal turbulence generation},
  author={Gao, Han and Han, Xu and Fan, Xiantao and Sun, Luning and Liu, Li-Ping and Duan, Lian and Wang, Jian-Xun},
  journal={Computer Methods in Applied Mechanics and Engineering},
  volume={427},
  pages={117023},
  year={2024},
  publisher={Elsevier}
}

@article{wei2024diffphycon,
  title={DiffPhyCon: A Generative Approach to Control Complex Physical Systems},
  author={Wei, Long and Hu, Peiyan and Feng, Ruiqi and Feng, Haodong and Du, Yixuan and Zhang, Tao and Wang, Rui and Wang, Yue and Ma, Zhi-Ming and Wu, Tailin},
  journal={Advances in Neural Information Processing Systems},
  volume={37},
  pages={4090--4147},
  year={2024}
}

@article{bodnar2025foundation,
  title={A foundation model for the Earth system},
  author={Bodnar, Cristian and Bruinsma, Wessel P and Lucic, Ana and Stanley, Megan and Allen, Anna and Brandstetter, Johannes and Garvan, Patrick and Riechert, Maik and Weyn, Jonathan A and Dong, Haiyu and others},
  journal={Nature},
  pages={1--8},
  year={2025},
  publisher={Nature Publishing Group UK London}
}

@article{lam2023learning,
  title={Learning skillful medium-range global weather forecasting},
  author={Lam, Remi and Sanchez-Gonzalez, Alvaro and Willson, Matthew and Wirnsberger, Peter and Fortunato, Meire and Alet, Ferran and Ravuri, Suman and Ewalds, Timo and Eaton-Rosen, Zach and Hu, Weihua and others},
  journal={Science},
  volume={382},
  number={6677},
  pages={1416--1421},
  year={2023},
  publisher={American Association for the Advancement of Science}
}

@article{bi2023accurate,
  title={Accurate medium-range global weather forecasting with 3D neural networks},
  author={Bi, Kaifeng and Xie, Lingxi and Zhang, Hengheng and Chen, Xin and Gu, Xiaotao and Tian, Qi},
  journal={Nature},
  volume={619},
  number={7970},
  pages={533--538},
  year={2023},
  publisher={Nature Publishing Group UK London}
}

@article{wang2025simulating,
  title={Simulating Three-dimensional Turbulence with Physics-informed Neural Networks},
  author={Wang, Sifan and Sankaran, Shyam and Stinis, Panos and Perdikaris, Paris},
  journal={arXiv preprint arXiv:2507.08972},
  year={2025}
}

@article{wang2024respecting,
  title={Respecting causality for training physics-informed neural networks},
  author={Wang, Sifan and Sankaran, Shyam and Perdikaris, Paris},
  journal={Computer Methods in Applied Mechanics and Engineering},
  volume={421},
  pages={116813},
  year={2024},
  publisher={Elsevier}
}

@article{sun2020surrogate,
  title={Surrogate modeling for fluid flows based on physics-constrained deep learning without simulation data},
  author={Sun, Luning and Gao, Han and Pan, Shaowu and Wang, Jian-Xun},
  journal={Computer Methods in Applied Mechanics and Engineering},
  volume={361},
  pages={112732},
  year={2020},
  publisher={Elsevier}
}

@article{raissi2019physics,
  title={Physics-informed neural networks: A deep learning framework for solving forward and inverse problems involving nonlinear partial differential equations},
  author={Raissi, Maziar and Perdikaris, Paris and Karniadakis, George E},
  journal={Journal of Computational physics},
  volume={378},
  pages={686--707},
  year={2019},
  publisher={Elsevier}
}

@article{goswami2020transfer,
  title={Transfer learning enhanced physics informed neural network for phase-field modeling of fracture},
  author={Goswami, Somdatta and Anitescu, Cosmin and Chakraborty, Souvik and Rabczuk, Timon},
  journal={Theoretical and Applied Fracture Mechanics},
  volume={106},
  pages={102447},
  year={2020},
  publisher={Elsevier}
}

@inproceedings{liu2022convnet,
  title={A convnet for the 2020s},
  author={Liu, Zhuang and Mao, Hanzi and Wu, Chao-Yuan and Feichtenhofer, Christoph and Darrell, Trevor and Xie, Saining},
  booktitle={Proceedings of the IEEE/CVF conference on computer vision and pattern recognition},
  pages={11976--11986},
  year={2022}
}

@article{dosovitskiy2020image,
  title={An image is worth 16x16 words: Transformers for image recognition at scale},
  author={Dosovitskiy, Alexey and Beyer, Lucas and Kolesnikov, Alexander and Weissenborn, Dirk and Zhai, Xiaohua and Unterthiner, Thomas and Dehghani, Mostafa and Minderer, Matthias and Heigold, Georg and Gelly, Sylvain and others},
  journal={arXiv preprint arXiv:2010.11929},
  year={2020}
}

@inproceedings{he2016deep,
  title={Deep residual learning for image recognition},
  author={He, Kaiming and Zhang, Xiangyu and Ren, Shaoqing and Sun, Jian},
  booktitle={Proceedings of the IEEE conference on computer vision and pattern recognition},
  pages={770--778},
  year={2016}
}

@article{loshchilov2017decoupled,
  title={Decoupled weight decay regularization},
  author={Loshchilov, Ilya and Hutter, Frank},
  journal={arXiv preprint arXiv:1711.05101},
  year={2017}
}

@article{Mota2021,
    title = {Effect of sub-boundaries on primary spacing dynamics during {3D} directional solidification conducted on {DECLIC}-{DSI}},
    volume = {204},
    issn = {13596454},
    url = {https://linkinghub.elsevier.com/retrieve/pii/S1359645420309253},
    doi = {10.1016/j.actamat.2020.116500},
    journal = {Acta Materialia},
    author = {Mota, F.L. and Pereda, J. and Ji, K. and Song, Y. and Trivedi, R. and Karma, A. and Bergeon, N.},
    month = feb,
    year = {2021},
    pages = {116500},
}

@article{jagtap2020extended,
  title={Extended physics-informed neural networks (XPINNs): A generalized space-time domain decomposition based deep learning framework for nonlinear partial differential equations},
  author={Jagtap, Ameya D and Karniadakis, George Em},
  journal={Communications in Computational Physics},
  volume={28},
  number={5},
  year={2020},
  publisher={Brown Univ., Providence, RI (United States)}
}

@article{li2019d3m,
  title={D3M: A deep domain decomposition method for partial differential equations},
  author={Li, Ke and Tang, Kejun and Wu, Tianfan and Liao, Qifeng},
  journal={Ieee Access},
  volume={8},
  pages={5283--5294},
  year={2019},
  publisher={IEEE}
}

@article{um2020solver,
  title={Solver-in-the-loop: Learning from differentiable physics to interact with iterative pde-solvers},
  author={Um, Kiwon and Brand, Robert and Fei, Yun Raymond and Holl, Philipp and Thuerey, Nils},
  journal={Advances in neural information processing systems},
  volume={33},
  pages={6111--6122},
  year={2020}
}

@article{fan2025diff,
  title={Diff-FlowFSI: A GPU-Optimized Differentiable CFD Platform for High-Fidelity Turbulence and FSI Simulations},
  author={Fan, Xiantao and Liu, Xinyang and Wang, Meng and Wang, Jian-Xun},
  journal={arXiv preprint arXiv:2505.23940},
  year={2025}
}

\end{document}